 \documentclass[preprint2]{aastex}

\slugcomment{Accepted for publicaiton in the AJ}

\shorttitle{Variability in PPNe: VI. Medium-bright, C-rich}

\shortauthors{Hrivnak et al.}

\begin{document}

\title{VARIABILITY IN PROTO-PLANETARY NEBULAE: VI. MULTI-TELESCOPE LIGHT CURVE STUDIES OF SEVERAL MEDIUM-BRIGHT (V=13$-$15), CARBON-RICH OBJECTS}

\author{Bruce J. Hrivnak\altaffilmark{1.5}, Gary Henson\altaffilmark{2,5}, Todd C. Hillwig\altaffilmark{1,5}, Wenxian Lu\altaffilmark{1}, Brian W. Murphy\altaffilmark{3,5}, and Ronald H. Kaitchuck\altaffilmark{4,5} }

\altaffiltext{1}{Department of Physics and Astronomy, Valparaiso University, 
Valparaiso, IN 46383; bruce.hrivnak@valpo.edu, todd.hillwig@valpo.edu, wen.lu@valpo.edu (retired)}
\altaffiltext{2}{Department of Physics and Astronomy, East Tennessee State University, 
Johnson City, TN 37614; hensong@mail.etsu.edu}
\altaffiltext{3}{Department of Physics and Astronomy, Butler University, Indianapolis, IN, USA; bmurphy@butler.edu}
\altaffiltext{4}{Department of Physics and Astronomy, Ball State University, Muncie, IN 47306; rkaitchu@bsu.edu}
\altaffiltext{5}{Southeastern Association for Research in Astronomy (SARA)}

\begin{abstract}
We present ten years of new photometric monitoring of the light variability of five evolved stars with strong mid-infrared emission from surrounding dust.  Three are known carbon-rich proto-planetary nebulae (PPNe) with F$-$G spectral types; the nature of the other two was previously unknown.     
For the three PPNe, we determine or refine the pulsation periods of IRAS 04296$+$3429 (71 days), 06530$-$0213 (80 days), and 23304+6147 (84 days).  A secondary period was found for each, with a period ratio {\it P$_2$}/{\it P$_1$} of 0.9.  The light variations are small, 0.1$-$0.2 mag.  These are similar to values found in other PPNe.  The other two are found to be giant stars.  IRAS 09296+1159 pulsates with a period of only 47 days but reaches pulsational light variations of 0.5 mag.  Supplemental spectroscopy reveals the spectrum of a CH carbon star.  IRAS 08359$-$1644 is a G1~III star that does not display pulsational variability; rather, it shows non-periodic decreases of brightness of up to 0.5 mag over this ten-year interval.  These drops in brightness are reminiscent of the light curves of R Corona Borealis variables, but with much smaller decreases in brightness, and are likely due to transient dust obscuration.  Its SED is very similar to that of the unusual oxygen-rich giant star HDE 233517, which possesses mid-infrared hydrocarbon emission features.  These two non-PPNe turn out to members of the rare group of giant stars with large mid-infrared excesses due to dust, objects which presumably have interesting evolutionary histories.
\end{abstract}


\section{INTRODUCTION}
\label{Intro}

Proto-planetary nebulae (PPNe) are objects in the short-lived evolutionary stage between asymptotic giant branch (AGB) stars and planetary nebulae (PNe).  They consist of a luminous, post-AGB star surrounded by a circumstellar nebula of gas and dust.  Most PPNe candidates were first identified based on the mid-infrared emission from their circumstellar dust as detected by the {\it Infrared Astronomical Satellite} ({\it IRAS}) and then visible candidates were identified for most of them through ground-based follow-up. 
Images showing both the central stars and the faint reflected light from the circumstellar dust have been obtained using the {\it Hubble Space Telescope} \citep{ueta00,sah07,siod08}.

Ground-based spectroscopic studies of the central stars have revealed that many of them have carbon-rich (C-rich) atmospheres \citep{vanwin03}.  This was initially seen by the presence of molecular carbon bands in their spectra \citep{hri95} and then followed up by high-resolution abundance analyses \citep{vanwin00,red02}.  These studies were of particular interest because they also revealed the presence of a high abundance of {\it s}-process elements, the expected product of neutron-capture during the AGB evolution of the stars.  Though less exciting from the standpoint of nucleosynthesis, abundance analyses of other PPNe have revealed the oxygen-rich (O-rich) nature of their atmospheres.  
Similarly, infrared spectroscopic analyses of these objects have revealed the dominant chemistry (C-rich or O-rich) of their circumstellar dust, which in almost all cases corresponds to that of their atmospheres.  

In this study, we focus on the light variability of the central stars of five objects with large mid-infrared excesses.  This is part of a series of papers in which we investigate variability in PPNe and related objects.  In previous studies, we have discussed the pulsational variability in 12 C-rich PPNe and four O-rich PPNe in the Milky Way Galaxy and 22 C-rich PPNe and PPN candidates in the Megallanic Clouds \citep{hri10,hri15b,hri15a}.  
Studies of pulsation in PPNe have also been published by Arkhipova and collaborators \citep[e.g.,][]{ark10,ark11}.
For the brighter ones ({\it V} $\le$ 10.5 mag), we have also compared their radial velocity curves with their light curves to determine the phase relationship between their brightness, temperature, and size \citep{hri13,hri18}.
Here we present our observed light curves for five medium-bright objects. 
Four of them display molecular carbon in their visual spectrum and are clearly carbon-rich; the fifth does not, but is found to have an unusual light curve.  
Their light curves are investigated and period analyses are carried out.  
We also analyzed recent, publicly-available light curves of these objects.  
These results are discussed in the context of previous studies of PPNe and related objects.

\section{PROGRAM OBJECTS}
\label{Objects}

The five objects in this program are listed in Table~\ref{object_list}. 
These objects are of medium brightness, with {\it V} = 13$-$15 mag.
In this table are listed their identifications in different catalogs, coordinates in the equatorial and galactic systems, {\it V} magnitudes, color indices, and spectral types.
The objects were chosen on the basis of their relatively large mid-infrared emission and, for three of them (IRAS 04296+3429, 065304-40213, and 23304+6147), their previous identification as PPNe.  IRAS 09296+1159 was previously observed to display some infrared emission features found in PPNe \citep{geb90}.
{\it Hubble Space Telescope} images have been taken of IRAS 04296+3429, 06530$-$0213, and 23304+6147, and they reveal a bipolar nebula in each at visible wavelengths \citep{ueta00,siod08}. 

\placetable{object_list}

\section{PHOTOMETRIC OBSERVATIONS AND REDUCTIONS}
\label{Obs}

Observations were carried out with three different telescopes.
The Valparaiso University Observatory (VUO) 0.4-m campus telescope was the main one used, and two of the program objects, IRAS 09296+1159 and 23304+6147, were only observed using it.  
Two telescopes operated by the Southeastern Association for Research in Astronomy \citep[SARA;][]{keel17} were also used: 
the 0.9-m at Kitt Peak National Observatory (SARA-KP), and  0.6-m Lowell telescope at Cerro-Tololo Interamerican Observatory (SARA-CT).  
Each was equipped with a CCD detector and standard Johnson {\it BV} and Cousins {\it RI} filters.  (The observations made through the latter will be designated {\it R$_C$} and {\it I$_C$}.) 
In this program, we are using the {\it V} and {\it R$_C$} filters.   
Each of the objects has been observed from the VUO, beginning in 2008 or 2009 and continuing through 2018, with the last observations made in May 2018.   
The other three objects were also observe with the SARA telescopes.  These observations were also begun in 2008 or 2009 and continued through April 2018.
However, for each of the SARA telescopes, we went through a succession of three different detectors and in one case two different filter sets. 
Typical exposure times at the VUO ranged from 8 to 30 min for the different objects.  The full-width, half-maximum (FWHM) of the VUO images were typically 2.5$-$3.0$\arcsec$.  The image quality was better and the exposure times were shorter at the SARA telescopes.

The observations were reduced using standard procedures in IRAF\footnote{IRAF is distributed by the National Optical Astronomical Observatory, operated by the Association for Universities for Research in Astronomy, Inc., under contract with the National Science Foundation.}.
This involved removing cosmic rays, subtracting the bias, and flat fielding the images. 
Aperture photometry was carried out, using an aperture of $\sim$11$\arcsec$ diameter.
While this was a larger aperture than is needed for the FWHM of the images, it is what was used in the reduction of earlier observations at the VUO with a different CCD \citep{hri10} and permitted easier combination with the earlier data into composite light curves.
The statistical uncertainty in the individual observations is $\le$ 0.010 mag.

Standard stars from the lists of \citet{land83,land92} were observed at each of the different telescope and detector systems, and the data were reduced using an aperture of 14$\arcsec$, as used by Landolt.  From these we have determined standard linear color coefficients that were used in the transformation from the instrumental to the standard photometric system.  
The standard magnitudes of the program stars are listed in Table~\ref{std_ppn}.
We carried out a program of differential photometry, using three comparison stars in each field.  The identifications of the comparison stars and their standard magnitudes are listed in  Table~\ref{std_comp}.

\placetable{std_ppn} 

\placetable{std_comp} 

The use of the different telescopes 
at different latitudes has allowed us to obtain more data than could be done with only one telescope.  
However, it also introduced some generally small systematic offsets into the data.
Of the three objects observed with the SARA telescopes, two are very red, with ({\it B$-$V)} $\ge$ 2.0, which is redder than an M5 star, and ({\it V$-$R$_C$}) $\ge$ 1.2, which is the color of an M0 giant or dwarf, and the other one has ({\it B$-$V)} = 1.2, the color of early K stars \citep{cox00}.
This extreme stellar redness, combined with the neglect of second-order effects in the color and extinction terms used in the transformation each telescope-filter-detector system to the standard system, is the likely cause of these systematic offsets.
Since all of the objects were observed from the VUO using a single detector and filter set, we have chosen to use the VUO data set as the reference system for each object and adjusted the other data to this system by the introduction of an offset.  
Offset values were determined primarily by determining the magnitude differences in data observed at both VUO and the other system on the same or adjacent nights. 
In a few cases where such direct comparison with the VUO data was not possible, we instead determined an offset by comparison with another data set for that star for which an offset from the VUO data was well established.
These were all checked visually to confirm that the determined offset appeared to improve the data compatibility. 
In the large majority of cases, the offset values were between $-$0.02 and $+$0.02 mag, but in a few cases they reached to $-$0.10 mag.
We estimate the uncertainty in these offset values to be $\pm$0.01$-$0.02 mag.  While this is larger than we would like, it should have little impact on the period determinations and only a small impact on the amplitude determinations.
More specific details and access to the data are listed in in Appendix I.

\begin{figure}\figurenum{1}\epsscale{1.10} 
\plotone{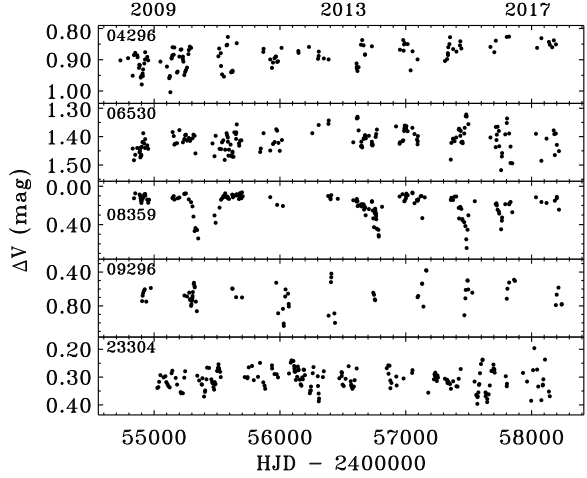}
\caption{Our new {\it V} light curves from 2008$-$2018.  (Typical error bars range from $\pm$0.010 to $\pm$0.020 mag and are shown in Figs. 6$-$8 and 10.)
\label{LC-new}}
\epsscale{1.0}
\end{figure}

The standardized, combined {\it V} light curves based on the recent 2008$-$2018 data are shown in 
Figure~\ref{LC-new}.
The uncertainty in the combined data is 0.01$-$0.025 mag, with a typical uncertainty of $\pm$0.015 mag.
It can be seen that all of the objects vary in light.  These light curves will be discuss individually and analyzed in Section~\ref{LC-Var}.

\placefigure{LC-new}

\begin{figure}\figurenum{2}\epsscale{1.10} 
\plotone{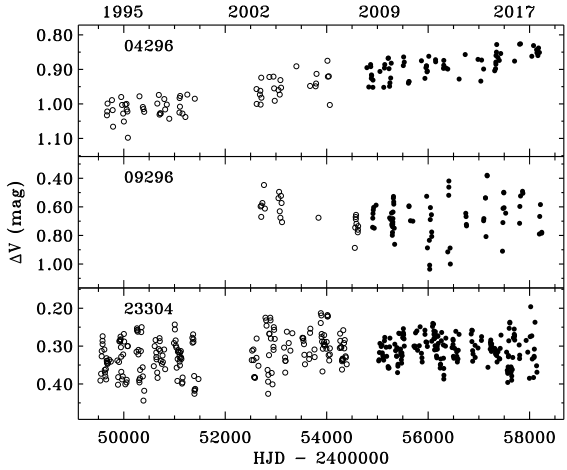}
\caption{Our older {\it V} light curves from 1994$-$2007 (open circles) combined with the newer ones from 2008$-$2018 (filled circles) to show longer-term trends in the data.
\label{LC-oldnew}}
\epsscale{1.0}
\end{figure}

 For three of the objects, IRAS 04296+3429, 09296+1159, and 23304+6147, we have earlier data obtained at the VUO with a different CCD detector which did not have auto-guiding capabilities.  This severely limited our integration times to $\le$ 6 min, with a consequential decrease in the precision of our observations.  
 An analysis of this earlier data for IRAS 04296+3429 and 23304+6147 was presented previously by \citet{hri10}; this is the first presentation of these data for IRAS 09296+1159.  
In Figure~\ref{LC-oldnew} are shown the earlier {\it V} light curve data for IRAS 04296+3429 and 23304+6147 from 1994$-$2007 and for IRAS 09296+1159  from 2003$-$2007, together with the recent data for these three objects.   These allow us to investigate the presence of longer-term trends in these three data sets.

\placefigure{LC-oldnew}

\begin{figure}\figurenum{3}\epsscale{0.90} 
\plotone{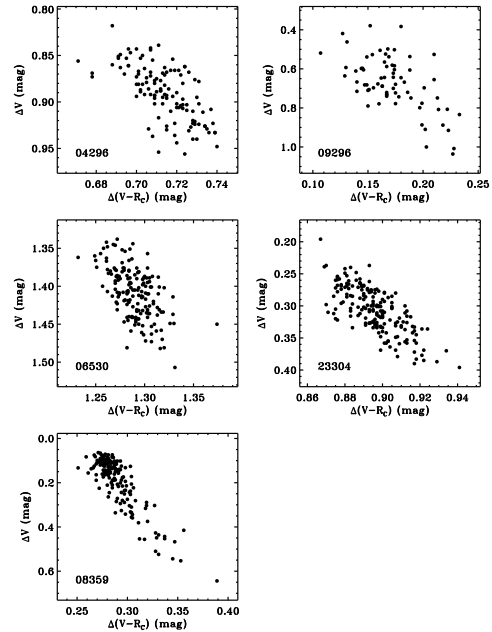}
\caption{The color curves for the target objects: ({\it V$-$R$_C$}) versus {\it V}.  All of them show a clear trend of increasing color with decreasing brightness.  The {\it V} light curve data for IRAS 04296+3429 and 06530$-$0213 have first had their long-term brightness trends removed.  
\label{color}}
\epsscale{1.0}
\end{figure}

The objects all vary in color.  This is shown in Figure~\ref{color}, which displays the {\it V} versus ({\it V$-$R$_C$}) color data derived from the 2008$-$2018 observations.  Each star is found to possess a general trend of being redder when fainter.  

\placefigure{color}

\section{ASAS-SN LIGHT CURVES}
\label{ASAS-SN}

\begin{figure}\figurenum{4}\epsscale{1.0} 
\plotone{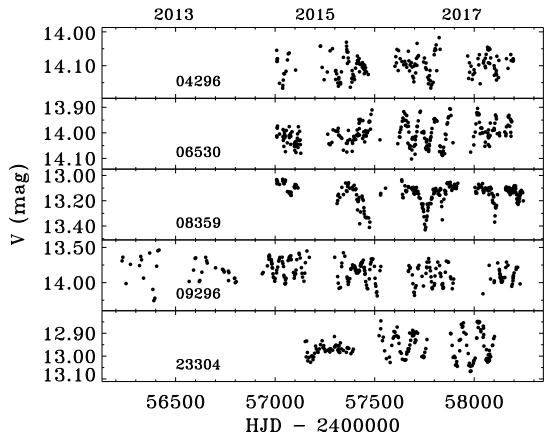}
\caption{The ASAS-SN {\it V} light curves from 2012$-$2018.
\label{LC-ASAS-SN}}
\epsscale{1.0}
\end{figure}

These target objects have also been observed with the automated telescopes of the All-Sky Automated Survey for Supernovae (ASAS-SN) \citep{koch17}, which began observations of IRAS 09296+1159 in late 2012 and the other four objects in late 2014 or 2015.  
We used observations concurrent with those that we had made, through April 2018.
These observations were made using cameras with 14 cm aperture telephoto lenses and CCD detectors, and they generally consist of three dithered observations of 90 sec each.  They were initially made only using a {\it V} filter.  The data are calibrated through observations of 100 nearby AAVSO Photometric All-Sky Survey objects \citep{hen12}.  Aperture photometry is used with a radius of 16$\arcsec$.
Since these telescopes are small in size and the exposure times are not long, the resulting data do not have the precision of our VUO and SARA data.  Additionally, ASAS-SN data sets cover fewer years. 
However, the observations are more numerous and denser in coverage, so we have used them as an additional source of information and will derive periods separately based on the ASAS-SN data.  
Only in the case of IRAS 08359$-$1644 have we combined the ASAS-SN data with our data; this will be discussed in Section~\ref{LC-Var}.
We first eliminated the data points with larger uncertainties.  
Since there were usually two or three successive observations made on a night, we averaged these and also formed an average standard deviation for the observation.
These light curves are shown in Figure~\ref{LC-ASAS-SN}.
All five of the program objects  have recently been recognized as variables in the ASAS-SN Catalog of Variable Stars.  

\placefigure{LC-ASAS-SN}

We also investigated data obtained earlier by the All Sky Automated Survey, version 3 (ASAS-3)\footnote{http://www.astrouw.edu.pl/asas/}, to see if it could be used to extend our knowledge of the light variations of these sources further back in time (2000$-$2009).  Data exist for the three southernmost objects, IRAS 06530$-$0213, 08359$-$1644, 09296+1159.  However, when we inspected the light curves, it was seen that the photometric uncertainties are too large (0.04$-$0.07) at these faint brightness levels to be useful for studying periodicity in these stars.

\section{LIGHT CURVES AND VARIABILITY STUDY}
\label{LC-Var}

\subsection{PERIODOGRAM ANALYSIS}
\label{periodogram}

Periodogram analyses on the light curve data were carried out using the program PERIOD04 \citep{lenz05}, which uses a Fourier analysis to search for significant frequencies.  This is a commonly used program, and it has the advantage of being able to fit one or more sine curves to the data.  
In an effort to confirm the reliability of the results from PERIOD04, we also analyzed the data using several other readily available periodogram programs, with similar results.  These supplemental analyses are discussed in Appendix II.
The reliability of the current results is also partially supported by the similar results determined from the independently observed ASAS-SN data, and for IRAS 04296+3429 and 23304+6147, the results of the analysis of the 1994$-$2007 data sets.
We used the standard criterion for significance of a signal-to-noise ratio (S/N) $\ge$ 4.0 \citep{bre93}, derived from a multi-site campaign to study pulsation in variable stars.

\subsection{IRAS 04296+3429}
\label{iras04296}

The light curves show variability superimposed on a general trend of increasing brightness, $-$0.07 mag in {\it V} and $-$0.06 mag in {\it R$_C$}, over the ten seasons of observations.  This continues the trend of increasing brightness seen in the observations from 1994$-$2007 of $-$0.10 mag in {\it V} over 13 years \citep{hri10} and shown in Figure~\ref{LC-oldnew}.  
Visual inspection of the recent data shows a cyclical variability of small amplitude.  
However, the amplitude of the cyclical variability from season to season (seasonal amplitude) changes.
It varies by approximately a factor of two over these ten recent seasons, with maximum values of 0.13 mag in {\it V} and 0.10 mag in {\it R$_C$}, with an amplitude ratio ({\it R$_C$}/{\it V}) of 0.7.  
The color varies over only a small range, $\Delta$(V$-$R$_C$) = 0.06 mag, and is slightly redder when fainter (Fig.~\ref{color}).

\begin{figure}\figurenum{6}\epsscale{1.0} 
\plotone{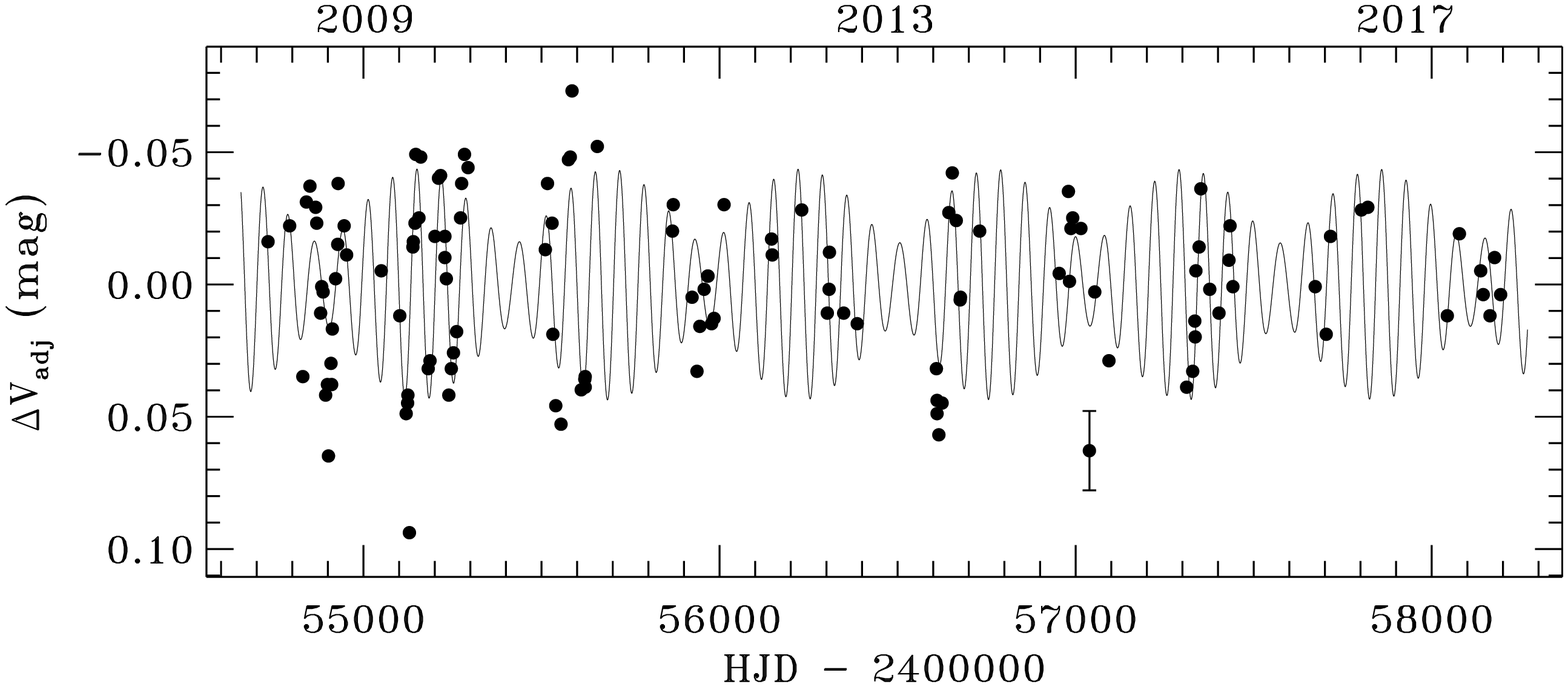}
\caption{Our trend-removed {\it V} light curve of IRAS 04296$+$3429 from 2008$-$2018, fitted by the two periods (71.3 and 63.0 days) and amplitudes listed in Table~\ref{periods}.
A typical error bar of $\pm$0.015 is shown on one data point.
\label{04296_LC}}
\epsscale{1.0}
\end{figure}

Analyses of the recent {\it V} and {\it R$_C$} light curves 
with the linear brightness trend removed result in similar dominant period values of 71.3$\pm0.1$ ({\it V}) and 71.2$\pm0.1$ ({\it R$_C$}) days.\footnote{Uncertainties quoted are statistical uncertainties determined by least-square fitting of the light curves in PERIOD04.  The time sampling of the varying light curves introduces additional uncertainties.} 
These confirm the more tentative period value of 71 days found from the earlier, lower-precision observations from 1994$-$2007.
A second, marginally significant period of 63.0$\pm0.1$ days is found in the larger amplitude {\it V} light curve.  
The analysis of the {\it V$-$R$_C$} color curve reveals a similar, but not formally significant (S/N$=$3.6), period of 71.4$\pm$0.2 days.
These period results are listed in Table~\ref{periods}, along with the associated amplitudes, phases, and standard deviations of the observations from the curve fits, and also the years over which the observations were made and the total number of observations in each filter.
The frequency spectrum for {\it P}$_1$ = 71.3 day and the resulting {\it V} phase curve are shown in Figure~\ref{Pspec}. 
In Figure~\ref{04296_LC} is shown the trend-removed {\it V} light curve fitted with the two periods, amplitudes, and phase values listed in the table.   One can see that these give a reasonably good fit to the light curve.  The standard deviation of a data point is 0.020 mag in {\it V}.
Clearly the observed light variation is more complicated than can be represented by the summation of two sine curves of constant amplitude.  Other studies of PPNe have shown similar deviations in the light curves and detailed spectroscopic studies have revealed the presence of shocks accompanying the pulsations \citep{leb96,zacs16}.

\placetable{periods} 

\placefigure{Pspec}

\placefigure{04296_LC}

We also examined the ASAS-SN {\it V} light curve, which covers four years but with many more data points.  This resulted in period values of 66.0$\pm$0.3 and 74.0$\pm$0.3 days.
However, the period fit for the 4th year was not as good as the others, and without it, based on the good fits to the 2014$-$2017 data, the period values were 73.7$\pm$0.4 and 67.6$\pm$0.5 days. 
These differ somewhat from the period values of 71.3 $\pm$ 0.1 and 63.0 $\pm$ 0.1 days determined from our 2008$-$2018 {\it V} light curves.

\subsection{IRAS 06530$-$0213}
\label{iras06530}

This object shows a gradual increase in mean brightness over the first eight seasons of $\sim$0.05 mag in {\it V} and $\sim$0.035 mag in {\it R$_C$}, then is approximately constant in mean brightness the last two seasons in {\it R$_C$} but with a slight decrease in the last two seasons of $\sim$0.02 mag in {\it V}. 
However, this apparent decrease in the last two seasons seems to not be real, since it is not seen in the much more numerous ASAS-SN {\it V} observations, as discussed below.  Likely it is the result of the small number of observations and the uncertainty in the offsets applied.
Superimposed on this trend is a variation in the seasonal amplitudes, 
which, for the years with the most data ($\ge$ 18 data points), vary over a range of a factor of three, from 0.05 mag (year 2) to 0.18 mag (year 9) in {\it V}.  The seasonal amplitude variations in {\it R$_C$} are smaller, typically 0.9 (0.87) those in {\it V}.
The color variation is small,  averaging $\Delta$({\it V$-$R$_C$}) = 0.055 mag in a season, and shows a trend of being redder when fainter (Fig.~\ref{color}).

\begin{figure}\figurenum{7}
\plotone{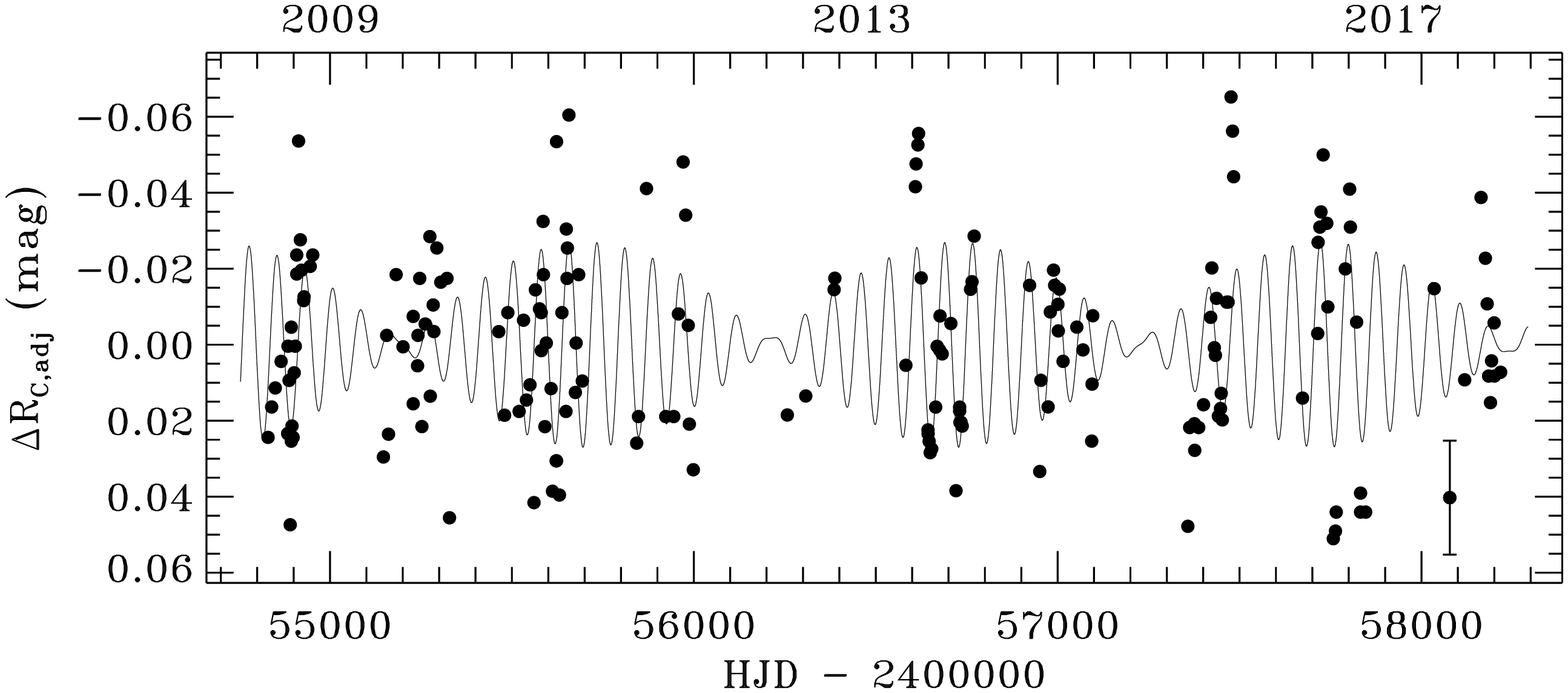}\epsscale{0.473} 
\plotone{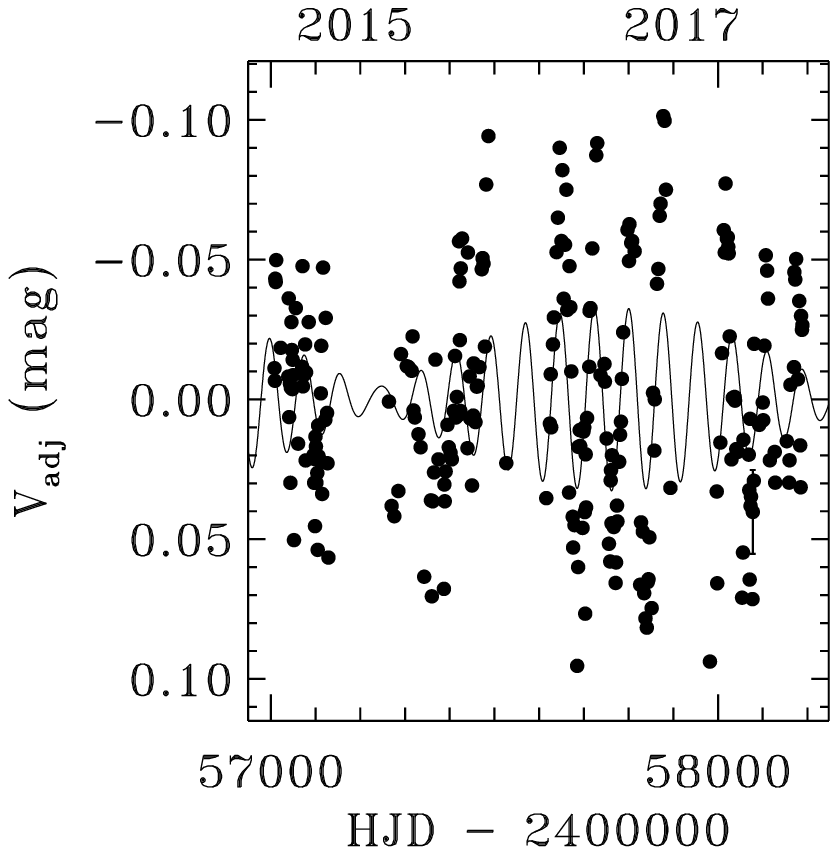}
\caption{(top) Our trend-removed, seasonally-normalized {\it R$_C$} light curve of IRAS 06530$-$0213 from 2008$-$2018, fitted by the two periods (79.5 and 73.7 days) and amplitudes listed in  Table~\ref{periods}.  A typical error bar of $\pm$0.015 is shown on one data point. 
(bottom) The ASAS-SN light curve from 2014$-$2018 fitted by the two-period fit based on 
our ten-year {\it V} light curve.
\label{06530_LC}}
\epsscale{1.0}
\end{figure}

For the period analyses, we first removed the trend in the light curves and then normalized the data to seasonal average values.
Two periods are found in each data set, a dominant period of 79.5$\pm$0.1 days and a not quite significant ({\it S/N}=3.9) secondary period of 73.9$\pm$0.2 days in {\it V}, and similar periods of 79.5$\pm$0.1 days and a second significant period of 73.7$\pm$0.1 days in {\it R$_C$}.  No period was found in the low-amplitude ({\it V$-$R$_C$}) color curve.
In Table~\ref{periods} are listed these results.
Since in the {\it R$_C$} light curve both periods were significant, we have chosen to display these data.
The frequency spectrum and resulting {\it R$_C$} phase curve for {\it P}$_1$ = 79.5 days are shown in Figure~\ref{Pspec}. 
The trend-removed, seasonally-normalized {\it R$_C$} light curve is shown in Figure~\ref{06530_LC}, fitted with the two significant periods.  
These periods give a reasonably good fit to the data, given the small amplitudes and scatter in the data. 

\placefigure{06530_LC}

An ASAS-SN {\it V} light curve exists for the past four seasons, 2014-2018, with a high density coverage.  It displays clear cyclical variations, particularly in the last two years (Fig.~\ref{LC-ASAS-SN}).  Periodogram analysis reveals two periods, 73.8$\pm$0.2 and 78.9$\pm$0.4 days.  These are the same two periods as found in the longer-term, ten-year light curve, to within 2 $\sigma$, but with a reversal in which period is the more dominant. 
The ASAS-SN light curve shows the same features as seen in the last four years of our observed light curves.  
To show this clearly, we also display in Figure~\ref{06530_LC} the ASAS-SN {\it V} light curve fitted with the parameters derived from our 2008$-$2018 {\it V} light curve.  

 \subsection{IRAS 08359$-$1644}
\label{iras08359}

\begin{figure}\figurenum{8}\epsscale{1.0} 
\plotone{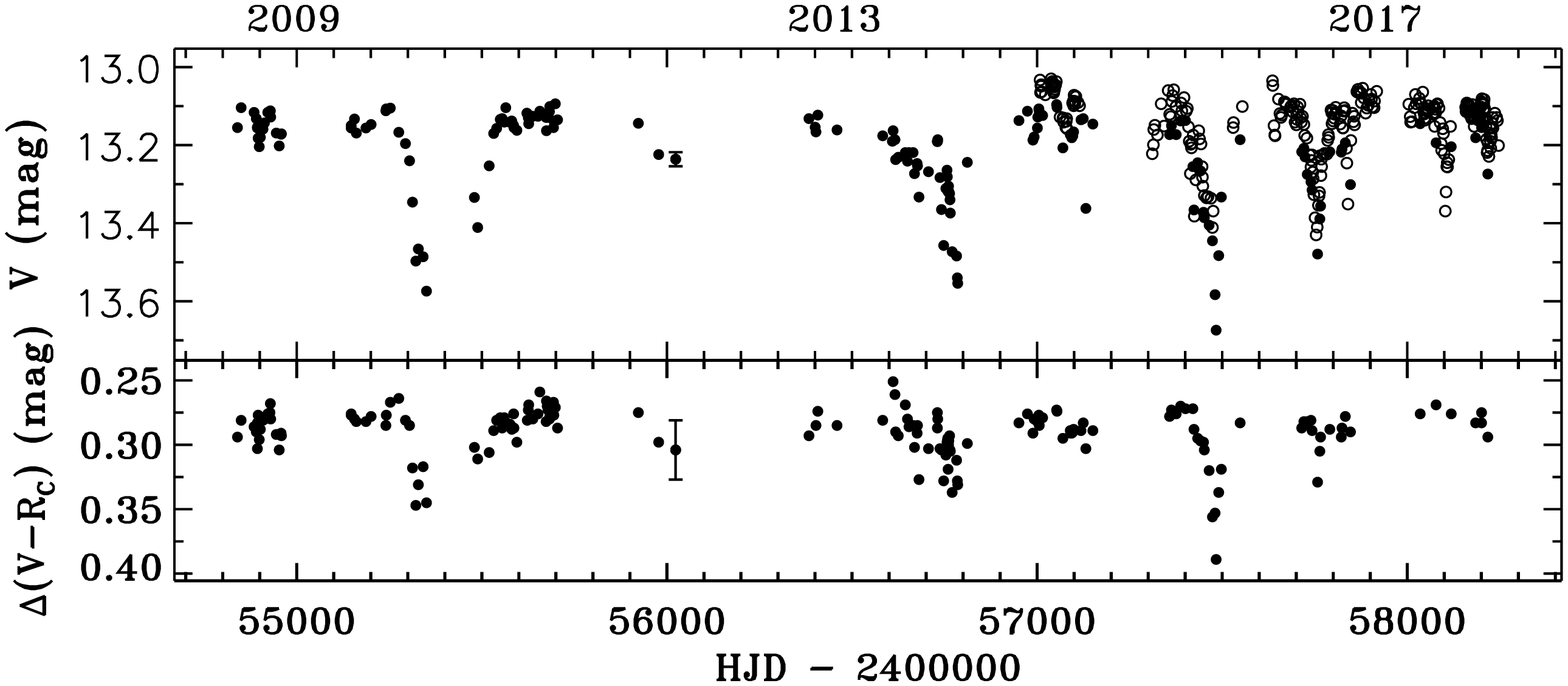}
\caption{(top) The combined {\it V} light curve of IRAS 08359$-$1644 from 2008$-$2018, based on our observations (filled circles) and the ASAS-SN data (open circles), with a magnitude offset added to convert from our differential to the standard magnitudes.
(bottom) Our observed differential $\Delta$({\it V$-$R$_C$}) color curve. 
Typical error bars from our observations 
of $\pm$0.018 ({\it V}) and $\pm$0.023 ({\it V$-$R$_C$}) are shown on one data point near the middle in each of the two panels.
\label{08359_LC}}
\epsscale{1.0}
\end{figure}

Our observations show variations at maximum light on the order of 0.1 mag (peak-to-peak) interspersed with decreases in brightness of 0.3$-$0.5 mag ({\it V}) over the ten seasons of observing.  The {\it R$_C$} light curve has a similar appearance but the decreases are not quite as large.  These variations are seen in even more detail over the past four seasons in the ASAS-SN {\it V} light curve.  In Figure~\ref{08359_LC} is shown the combined {\it V} light curve.  This was formed by adding to our observed differential light curve an offset determined from four nights observed in common on the two systems. 
Minima with decreases of $>$0.3 mag ({\it V}) are seen on 2,455,350,  2,456,785, 2,457,483, and 2,457,753, and there are several others ($\sim$2,455,489, $\sim$2,457,131, $\sim$2,458,102) with drops of $\sim$0.2 mag.
Inspection of the lower-level variations around maximum light suggest a possible cyclical variation in season 1 of about 40$-$50 days and an amplitude of $\sim$0.1 mag (peak-to-peak) and less certain cyclical variations of $\sim$100 days and amplitudes of $\sim$0.06 mag in seasons 2 and 3, but for the last of these seasons this cyclical variation is not maintained.  Season 6 also shows a cyclical variation, of $\sim$60 days with an amplitude of $\sim$0.1 mag, and in
seasons 8, 9, and 10, there are suggestions of possibly cyclical variations of $\sim$35$-$40 days and amplitudes of 0.06$-$0.1 mag.
The overall {\it V} light curve variation is 0.58 mag.
The color gets redder at the larger drops in brightness.  During the four decreases in brightness listed above, increases in ({\it V$-$R$_C$}) are seen ranging from 0.05 to 0.10 mag.  These are also shown in Figure~\ref{08359_LC}.

\placefigure{08359_LC}

A periodogram analysis of the combined {\it V} light curve yields a formally significant period of 164$\pm$1 days.  
However, when the data are plotted on a phase curve, the resulting light curve does not show good agreement among the drops in the light curve, which are scattered over a range of 0.35 in phase.  
Rather, these brightness decreases do not appear to be periodic.  
The cause of these relatively sudden drops is discussed in Section~\ref{discus}.

 \subsection{IRAS 09296$+$1159}
\label{iras09296}

This object was observed only at the VUO and the light curves are rather sparse, with a total of only 73 and 84 data points in {\it V} and {\it R$_C$}, respectively, and only 2 ({\it V}) or 3 ({\it R$_C$})  seasons with 10 or more observations.  They vary over a relatively large range, peak to peak, of 0.62 ({\it V}) and 0.52 ({\it R$_C$}) mag.
Visual inspection of the light curves in individual seasons reveals cyclical variations. 
The seasonal amplitudes vary quite a bit, although some of the seasons have only a few observations.  They reach maximum values of 0.58 ({\it V}) and 0.51 ({\it R$_C$}) mag in 2012 (season 5), with an average seasonal ratio ({\it R$_C$}/{\it V}) of 0.92.
The average seasonal values are also seen to vary (Fig.~\ref{LC-new}).  If we compare the median values of the extrema on the seasons where one can more clearly see the cyclical behavior, we find a range in the median seasonal values of 0.18 ({\it V}) and 0.15 ({\it R$_C$}) mag.  This variation in the average seasonal means is also seen in the less-precise, older VUO data (Fig.~\ref{LC-oldnew}) from 2003, 2004, and 2007.
The color varies over a relatively large range of 0.12 mag.  This object shows a very clear trend of color with brightness, being fainter when redder, as shown in Figure~\ref{color}.  

The VUO data sets are not well suited for a robust analysis due to the facts that the seasonal median values change by a relatively large amount and the data sets are small in number.  These make it difficult to normalize the individual seasonal values.  
We carried out our period analysis in two ways, directly on the observed data and also on a subset of five seasons (2, 4, 5, 7, 8) in which the extrema in the cyclical light curves were relatively well defined and from which we could get median values to use to normalize the seasonal data sets.  
The results were similar in most cases, with a dominant period of 40.2$\pm$0.1 days 
and a secondary period of 44.2$\pm$0.1 days in both {\it V} and {\it R$_C$}.  
The only exception was the normalized {\it R$_C$} data set, for which we found a secondary period of 54.5$\pm$0.1 days.
Examining the ({\it V$-$R$_C$}) data in a similar way, we found periods of 46.1 and 40.2 days for the normalized data and a not quite significant (S/N=3.7) period of 46.0 days for the full observed data set.

\begin{figure}\figurenum{9}\epsscale{1.0} 
\plotone{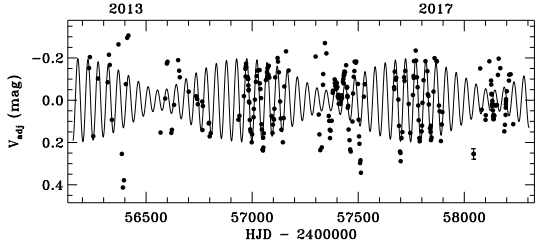}
\caption{The ASAS-SN {\it V} light curve of IRAS 09296$+$1159 from 2012$-$2018, fitted by the two periods (46.8 and 49.8 days) and amplitudes listed in  Table~\ref{periods}. 
A typical error bar of $\pm$0.025 is shown on one data point.
\label{09296_ASAS-SN}}
\epsscale{1.0}
\end{figure}

Fortunately the ASAS-SN {\it V} light curve, by contrast, is well populated over six seasons, 2012$-$2013 to 2017$-$2018, with over 50 points in each of seasons 3, 4, and 5.  They each show a cyclical pattern with a period of $\sim$50 days, but with seasonal amplitudes that vary by a factor of two, ranging from 0.72 mag in 2012$-$2013 to 0.36 mag in 2013$-$2014.  The seasonal light curves change in their average light levels over a range of 0.13 mag.  
An analysis of the observed light curve reveals a dominant period of 46.8$\pm$0.1 days, with a secondary period of 49.8$\pm$0.1 days.  
Similar values are found when we first normalize the light curves to the same seasonal mean values.  The dominant period gives a reasonably good fit to the periodicity in the light curve, with the secondary period primarily serving to modulate the amplitude.  However, under closer inspection, one can see that in some of the seasons the fits appears to be slightly off in phase.  We explored this further by analyzing the individual light curves in the later four seasons, which have the largest number of individual observations.  These yielded seasonal periods of 48.3, 47.3, 47.2, and 47.3 days for the seasons 2014$-$2015 to 2017$-$2018, respectively.  While the last three seasons have approximately the same period, the associated phases are different, such that an analysis of the combined light curve from these three seasons does not give a good fit to each season and actually results in a period of 48.2 days.  The results of the period analysis of the combined light curve for the six seasons is listed in Table~\ref{periods} and the fit to the light curve is shown in Figure~\ref{09296_ASAS-SN}.  

For comparison, we analyzed the VUO {\it V} light curve over the same six-season time interval (2012-2018), but using only the three seasons for which we could normalize the data sets.  This resulted in only 20 data points.  However, they resulted in in a good fit for two periods, with a dominant period of 46.7$\pm$0.1 days and a secondary period of 42.5$\pm$0.1 days.  Thus we find the same dominant period as found in the more numerous ASAS-SN data set over the same time interval.   
We note that an analysis of the Northern Sky Variability Survey data for this object (NSVS 10229563) resulted in a similar period of 46.9 days \citep{shin12}, based on observations made in 1999$-$2000 at a much lower precision than ours.  

\placefigure{09296_ASAS-SN}

 \subsection{IRAS 23304+6147}
\label{iras23304}

We previously observed this object from 1994$-$2007, at lower precision, and determined a dominant period of 85 days, with a secondary period of $\sim$71 days \citep{hri10}.  These earlier observations showed a slight increase in seasonal brightness of $\sim$0.04 mag in {\it V} and {\it R$_C$} over 13.5 years (see Fig.~\ref{LC-oldnew}). 

The recent observations from 2009$-$2018 do not show a systematic trend in seasonal brightness, but possess small seasonal variations in average brightness on the order of $\pm$0.01 mag.  The individual seasons show generally cyclical variations with changing amplitudes, with seasonal peak-to-peak values ranging from 0.08 to 0.19 mag in {\it V} and from 0.06 to 0.14 mag in {\it R$_C$}, with an average {\it R$_C$}/{\it V} amplitude ratio of 0.7.  The last two seasons show the largest variations in brightness.
The ({\it V$-$R$_C$}) values vary over a total range of 0.07 mag.

\begin{figure}\figurenum{10}\epsscale{1.0} 
\plotone{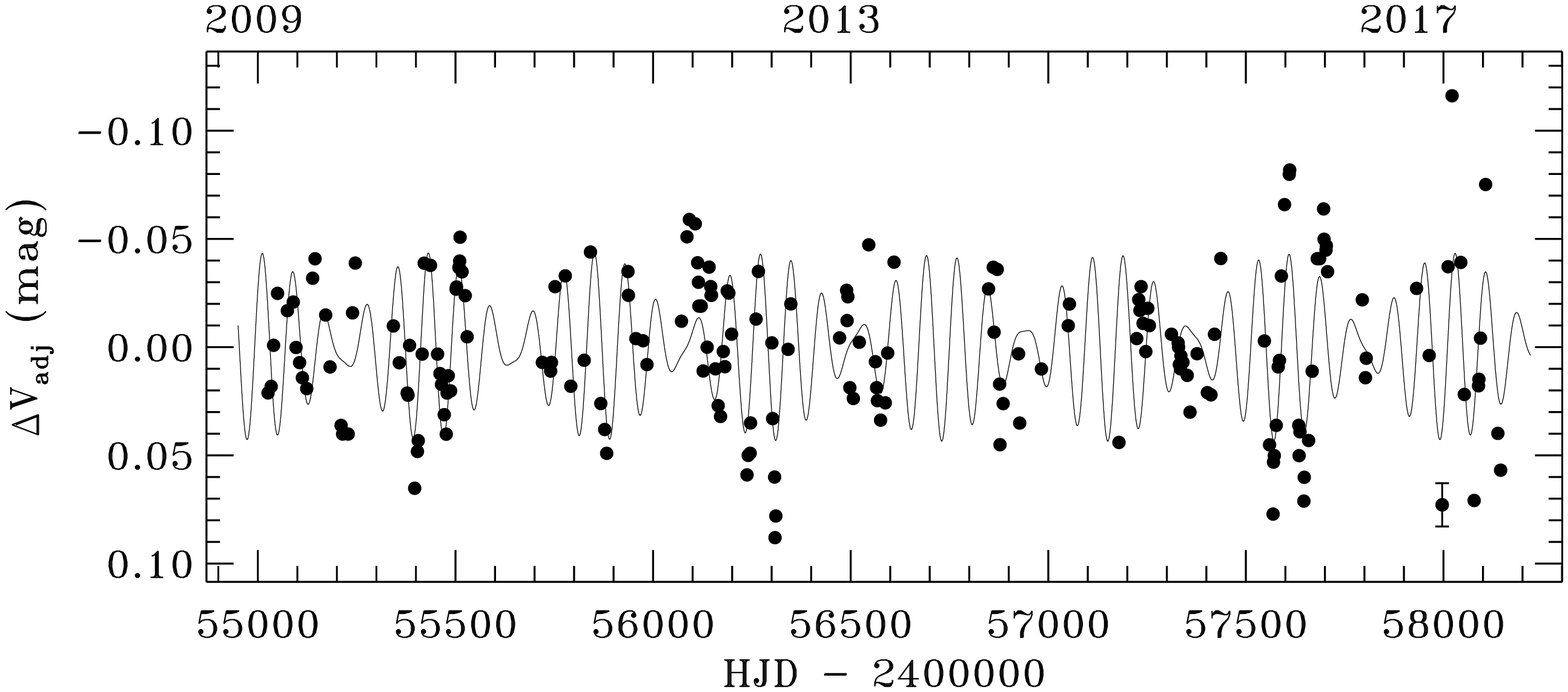}
\caption{Our seasonally-normalized {\it V} light curve of IRAS 23304$+$6147 from 2009$-$2018, fitted by the two periods (83.8 and 70.2 days) and amplitudes listed in  Table~\ref{periods}.
 A typical error bar of $\pm$0.010 is shown on one data point.
\label{23304_LC}}
\epsscale{1.0}
\end{figure}

We carried out a period analysis on each of the {\it V}, {\it R$_C$}, and ({\it V$-$R$_C$}) light curves.  These were done using first the observed light curves and then the normalized seasonal light curves, with very similar results.
The {\it V} light curve shows a dominant period of 83.8$\pm$0.1 days and a second significant period 70.2$\pm$0.1 days, with similar values seen in the {\it R$_C$} and ({\it V$-$R$_C$}) light curves.  The results are listed in Table~\ref{periods} and confirm the results found earlier in our lower-precision data.
The frequency spectrum for the {\it V} light curve and the associated phase plot based on {\it P$_1$} are shown in Figure~\ref{Pspec}.
The fits of the model sine curves with two periods to the seasonally-normalized light and color curves are of fair agreement, as shown in Figure~\ref{23304_LC}, with a standard deviation of 0.026 mag in {\it V}.  
However, in IRAS 23304+6147, as in the other stars in this study, there is clearly more variability in the star than can be accounted for by pulsation with two periods of fixed amplitudes.
An analysis of the last two seasons, which possess the largest amplitudes and clear cyclical variability, reveals a dominant period of 84.2$\pm$0.2 days in {\it V}, with similar values in {\it R$_C$}, and {\it V$-$R$_C$}, and very good fits to the light and color curves.  
 
\placefigure{23304_LC}

We also analyzed the ASAS-SN data set, which covers three seasons, beginning in 2015.  The light curve is similar to that obtained at the VUO for the same three years, in that the 2015-2016 season has a low amplitude and not a clear cyclical pattern while the 2016-2017 and 2017-2018 data show very clear cyclical patterns with increasingly large amplitudes.  The analysis of the three seasons of data results in two periods of 81.6$\pm$0.3 and 87.6$\pm$0.4 days, with the two periods producing a beat phenomenon that damps out the combined amplitude in the first season but reinforces it in the third.  When we analyze only the last two seasons, we find a period of 84.0$\pm$0.3 days, the same period found in the VUO data over the same time interval.  

\section{DISCUSSION}
\label{discus}

Four of the objects clearly display periodic variability in their light curves, while the other one (IRAS 08359$-$1644) shows several non-periodic drops in its brightness over the 10 years of observations.  
To better interpret these light curves and the nature of these objects, we begin with a discussion of their spectral energy distributions (SEDs) and their visible spectra.

\begin{figure}\figurenum{11}\epsscale{1.0} 
\plotone{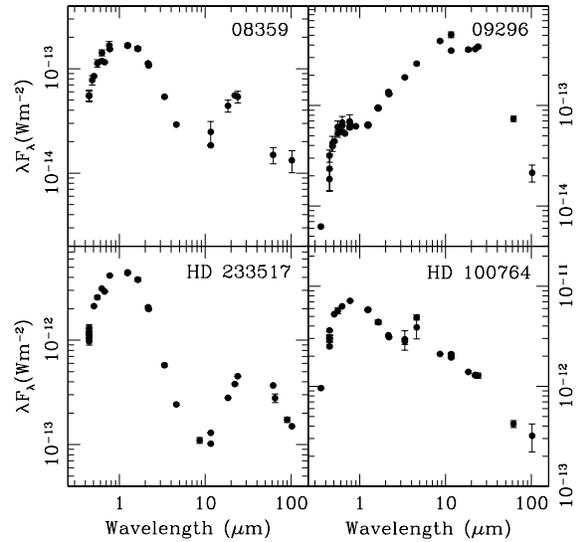}
\caption{Plots of the spectral energy distributions of IRAS 08359$-$1644 and 09296+1159, compared with relatively similar ones of the giant stars HD 233517 and HD 100764. 
Plotted are the {\it IRAS}, {\it 2MASS}, {\it WISE}, {\it Akari}, Johnson {\it BVJHK}, {\it Gaia}, and Sloan Digital Sky Survey data, with error bars, 
as listed in SIMBAD-VizieR.  In many cases the error bars are too small to see on this scale.
\label{SEDs}}
\epsscale{1.0}
\end{figure}

Four of the objects show a clearly bimodal SED.  This is common for PPNe, with one peak in the visible/near-infrared arising from the (reddened) photosphere and the second in the mid-infrared arising from the cool dust \citep[e.g.,][]{hri89}.  SEDs have been published for IRAS 04296+3429, 06530$-$0213, 23304+6147 \citep{kwo89, ueta00}.  
They each show a rising flux with increasing wavelength up to a peak in the {\it H} band (1.6 $\mu$m), then a decrease down to about 5 $\mu$m at which point the mid-IR component begins to dominate and the flux rises up to peak at about 25 $\mu$m before again decreasing.  In these three, the energy received in the mid-IR is 5$-$6 times that in the visible.    
The SEDs for IRAS 08359$-$1644 and 09296+1159 are shown in Figure~\ref{SEDs}, and they differ from those of these three PPNe.  
IRAS 08359$-$1644 also displays a clearly bimodal SED, but in this case the visible peak is 3 times stronger than the mid-IR peak and peaks at a slightly shorter wavelength (1.0 $\mu$m) than seen in the three PPNe.  
For IRAS 09296+1159, the SED shows two peaks but they are not so distinct.  The photospheric emission appears to peak at $\sim$0.7 $\mu$m and the much stronger dust emission peaks at $\sim$12 $\mu$m.  However, the mid-infrared peak is much broader than that seen in the other four objects, with the rise from the photospheric to the dust emission beginning at about 1 $\mu$m.

\placefigure{SEDs}

Visible-band spectroscopic observations have been published for three of the objects, IRAS 04296+3429, 06530$-$0213, and 23304+6147, beginning with low-resolution classification spectra \citep{hri95}.  They display the spectra of F$-$G supergiants but with molecular carbon absorption features (C$_2$ and C$_3$).
Photospheric abundance studies carried out using high-resolution spectra of IRAS 04296+3429, 06530$-$0213, and 23304+6147 confirm the high carbon abundance \citep{vanwin00,rey04}.  They also document the enhanced abundance of {\it s}-process elements, helping to confirm the PPN nature of these three objects.

\begin{figure}\figurenum{12}\epsscale{1.0} 
\plotone{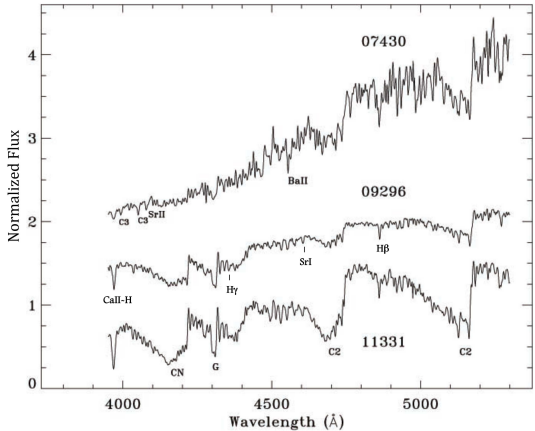}
\caption{The visible spectrum of IRAS 09296+1159, compared with that of the carbon-rich PPN IRAS 07430+1115 and the carbon star HD 100764 (IRAS 11331$-$1418).
(The spectra have been offset vertically for comparison.)
\label{09296_spec}}
\epsscale{1.0}
\end{figure}

\begin{figure}\figurenum{13}\epsscale{0.8} 
\rotatebox{270}{\plotone{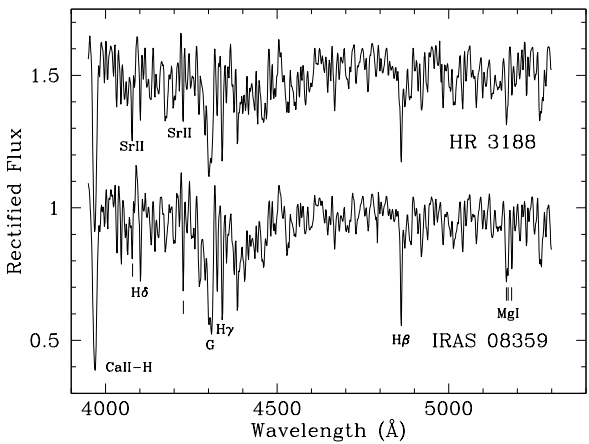}}
\caption{The visible spectrum of IRAS 08359$-$1644, compared with the G2~Ia standard HR 3188.  (The spectra have been offset vertically for comparison.)
\label{08359_spec}}
\epsscale{1.0}
\end{figure}

However, for IRAS 09296+1159 and 08359$-$1644, spectroscopic observations have not been published.  
We obtained spectra of these two objects in the course of an observing program to determine the spectral types of PPNe and PPN candidates, which will be discussed in detail elsewhere.  
The observations of these particular objects were carried out using the 2.3 m telescope and Boller and Chivens spectrograph at the Steward Observatory on 2000 Jan 04 and 05 (UT).  
The spectra have a resolution ($\Delta$$\lambda$) of 3.5 \AA~ and the reduced spectra range from 3950 to 5300 \AA.
The spectrum of IRAS 09296+1159 is shown in Figure~\ref{09296_spec}, where it is compared with the carbon-rich PPN IRAS 07430+1115 (G5~0-Ia \citep{hri99}; {\it B$-$V}=1.86, {\it V$-$R$_C$}=0.99; T=6000 K \citep{red02}) and the carbon star HD 100764 (IRAS 11331$-$1418; C-H2.5 \citep{kee93}; {\it B$-$V}=1.02, {\it V$-$R$_C$}=0.71; T=4750 K \citep{gos16}),
which were observed on the same nights. 
One can immediately see the carbon bands in IRAS 09296+1159.  
The hydrogen lines are strong, and barium and strontium lines are clearly present, although not nearly as strong as in IRAS 07430+1115.  The metal lines are relatively weak.
IRAS 09296+1159 also has a large space velocity, as discussed later in this section.
We thus classify it as a CH carbon star, similar in temperature to HD 100764, but with weaker C$_2$ bands.
This is in agreement with the recent classification of IRAS 09296+1159 as a CH carbon star on the basis of observations made with the Large Sky Area Multi-Object Spectroscopic Telescope \citep[LAMOST;][]{ji16}, covering the wavelength range 3800$-$7000 \AA~ at a resolution of $\Delta$$\lambda$/$\lambda$ = 1800.
The carbon bands are also seen in the PPNe IRAS 07430+1115, but lie on a much-more reddened continuum.
The spectrum of IRAS 08359$-$1644 is shown in Figure~\ref{08359_spec}.
Based on comparison of metal lines with the hydrogen lines, we determine a spectral classification of early-G.  
The luminosity class appears to be that of a giant, based on the strength of the CN band at $\lambda$4215 and ratios such as \ion{Y}{2} 3983/4005, \ion{Sr}{2} 4077/H$\delta$,  \ion{Fe}{1} 4046,4064, and \ion{Sr}{2} 4216/Fe 4144. 
In Figure~\ref{08359_spec} it is compared to the G2~Ia standard HR 3188.  IRAS 08359$-$1644 appears to be slightly hotter based on the hydrogen lines.  Therefore we classify it as G1~III.
Thus it appears, on the basis of their spectra, that three of the five objects are carbon-rich PPNe, one is a carbon star, and one is an evolved, probably oxygen-rich, giant.  

\placefigure{09296_spec}  
\placefigure{08359_spec} 

Pulsation periods have been determined for the three PPNe in this study.  For IRAS 06530$-$0213,  this is the first determination of its period, while for IRAS 04296+3429 and 23304+6147, these periods represent similar but better-determined values based on more precise observations.
All three also have secondary periods, which in each case have values slightly less than the primary periods.  These ratios, {\it P$_2$}/{\it P$_1$}, range from 0.84 to 0.93.  These values are similar to the range of period ratios of 0.81 to 1.16 that we found previously for twelve C-rich and four O-rich PPNe in the Milky Way Galaxy and eight C-rich PPNe in the Large and Small Magellanic Clouds \citep{hri13,hri15b,hri15a}. 

Previous studies of several bright PPNe \citep{hri13,hri18} have shown a close phase relationship between brightness and color, with all of the stars faintest when they are reddest.  
This relationship was investigated for these three PPNe.  We begin by adopting the ephemeris\footnote{The ephemeris is based on the arbitrary epoch of 2,544,600.0000 and the period value determined from the fit to the {\it V} data set.} from the {\it V} light curve and applying it to the {\it R$_C$} and ({\it V$-$R$_C$}) data sets for each star.  From this, we find that for each of these PPNe, minimum light in the {\it R$_C$} and ({\it V$-$R$_C$}) light curves occurs at the same phase ($\phi$) as in the {\it V} light curve, to within the uncertainty of the data.
For IRAS 04296+3429, we find that $\Delta$$\phi$({\it R$_C$}) = $\phi$({\it R$_C$}) $-$ $\phi$({\it V}) = 0.02 $\pm$ 0.02 and $\Delta$$\phi$({\it V$-$R$_C$}) = $\phi$({\it V$-$R$_C$}) $-$ $\phi$({\it V}) = $-$0.02 $\pm$ 0.04.
Similarly for IRAS 23304+6147, $\Delta$$\phi$({\it R$_C$}) = $-$0.01 $\pm$ 0.02 and $\Delta$$\phi$({\it V$-$R$_C$}) = 0.02 $\pm$ 0.03.
For IRAS 06530$-$0213 this is also the case, even though an analysis of the ({\it V$-$R$_C$}) data does not show a periodicity: $\Delta$$\phi$({\it R$_C$}) = 0.00 $\pm$ 0.03 and $\Delta$$\phi$({\it V$-$R$_C$}) = 0.02 $\pm$ 0.08.
A similar phase relationship is also seen for the carbon star IRAS 09296+1159.  Basing the analysis on the five seasons for which we could reasonably normalize the VUO data for this star, we found that $\Delta$$\phi$({\it R$_C$}) = 0.00 $\pm$ 0.04 and $\Delta$$\phi$({\it V$-$R$_C$}) = 0.04 $\pm$ 0.04.
Thus we find that all three of these PPNe are reddest, and thus coolest, at the same phase in their pulsation cycle as when they are faintest, in agreement with that found previously for several other well-studied PPNe.

\begin{figure}\figurenum{14}\epsscale{1.0} 
\plotone{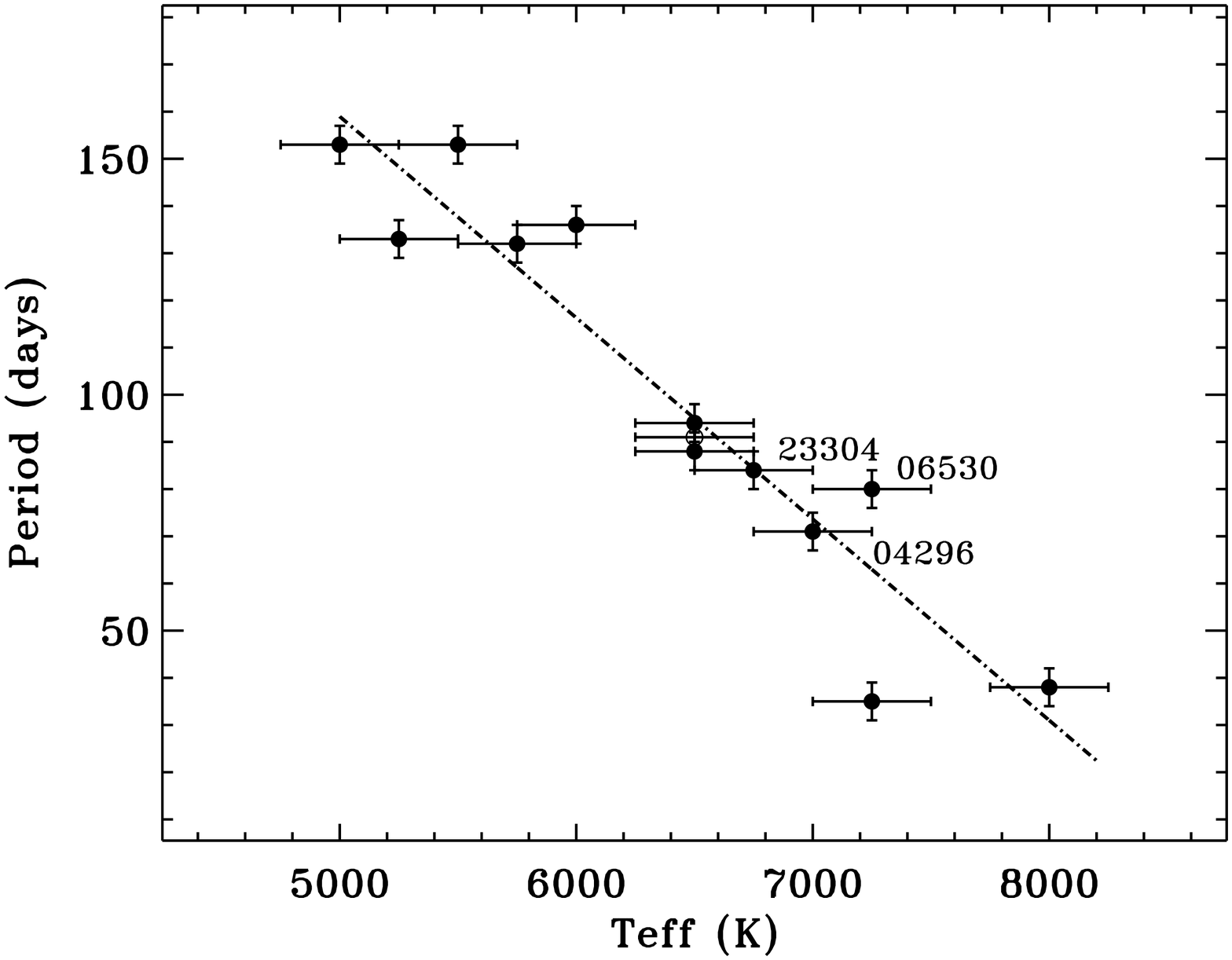}
\plotone{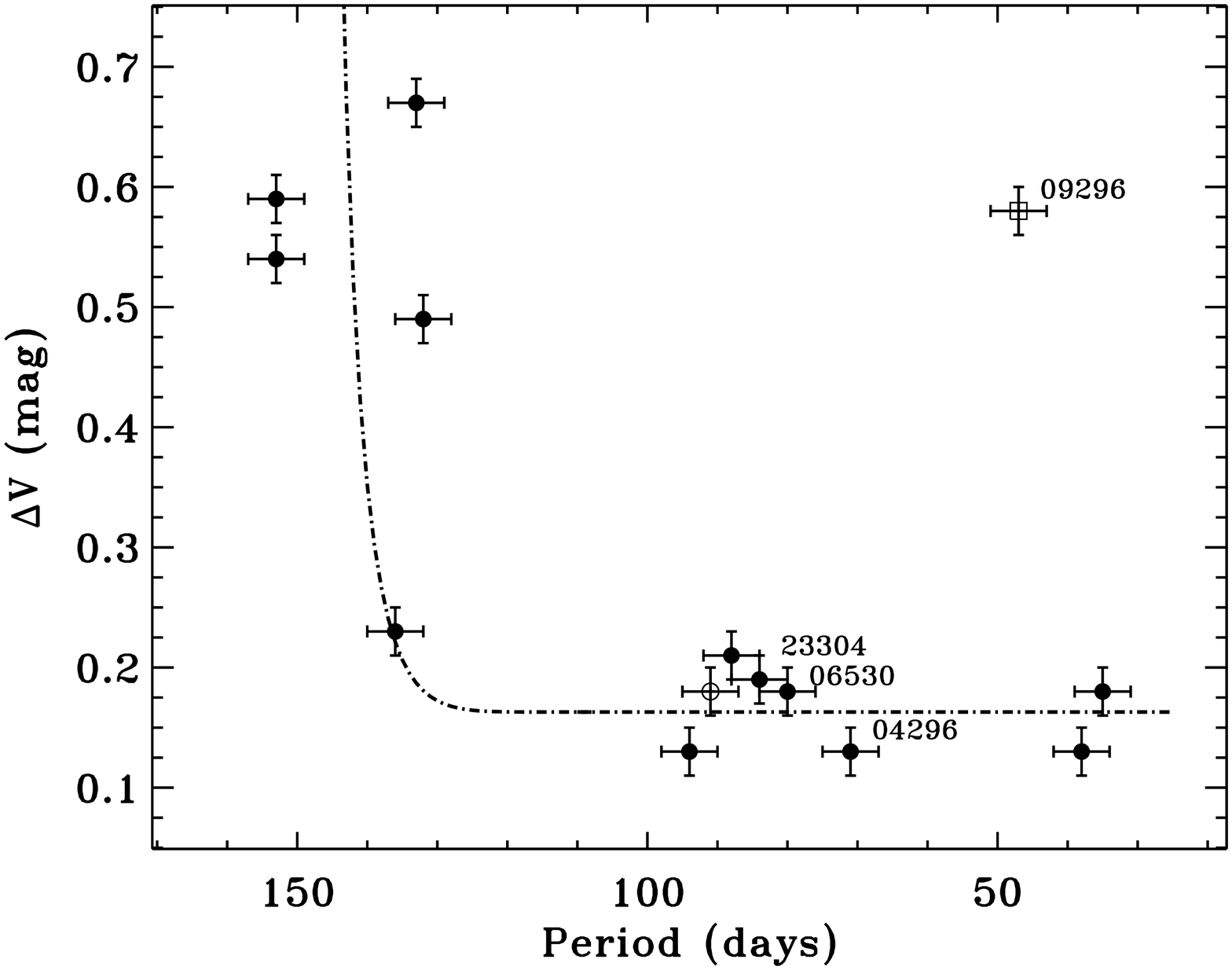}
\caption{Plots comparing the pulsation properties and temperatures of these three PPNe (labeled) with other carbon-rich PPNe from the study by \citet{hri10}.  The three PPNe from this study are labeled.
(top) Plot of primary pulsation periods versus effective temperatures of carbon-rich PPNe.  Error bars represent the estimated uncertainties of $\pm$250 K and the very conservative estimate of $\pm$3 days.   The data display a linear relationship of decreasing period with increasing temperature.  
(bottom) Plot of maximum variation in a season ($\Delta${\it V}) versus the primary pulsation period for carbon-rich PPNe.  The data indicate much larger seasonal variations at longer periods and an approximately constant low level of variation at periods $<$100 days.  Error bars represent the estimated uncertainties of $\pm$3 days (very conservative) and $\pm$0.02 mag.   The line is simply a free-hand representation of the trend.  
Also included is the pulsating carbon star IRAS 09296+1159 (open square).
\label{P-T}}
\epsscale{1.0}
\end{figure}

We previously found that there exists a relationship between the pulsation period, photospheric temperature, and pulsation amplitude for PPNe of F$-$G spectral types \citep{hri10}.  
In Figure~\ref{P-T} are shown some of these relationships, and it can be seen that IRAS 06530$-$0213 fits well with these, as do IRAS 04296+3429 and 23304+6147, which were part of the original study.  
The primary pulsation period shows a linear decrease with increasing temperature over the temperature range of these stars, 5000 to 8000 K.
The maximum seasonal variation in light, on the other hand, shows a highly non-linear trend, 
with relatively large (0.45$-$0.70 mag) values for four of the five PPNe with periods $>$130 days, but much smaller (0.12$-$0.22 mag) values for P $<$100 days.  IRAS 06530$-$0213 has a small maximum variation and fits in well with the others.  

\placefigure{P-T}

The carbon star IRAS 09296+1159 also shows periodic variability, with a dominant period of 46.8 days, which is shorter than those of most PPNe (see Fig.~\ref{P-T}, where it differs greatly from the P-T$_{eff}$ trend seen for PPNe).  
This short period is surprising, however, since if it were an AGB carbon star, it would be expected to have a much longer period, typically in the range of 200 to 700 days \citep{white06}. 
This shorter period suggests that IRAS 09296+1159 is smaller in size than an AGB carbon star.  
CH stars, which are giants, and dwarf carbon stars are both thought to be extrinsic carbon stars, which have obtained their enhanced carbon from the transfer of carbon-rich material from an evolved AGB carbon star companion onto their atmospheres \citep{mcc90,gre13}.  
This is in contrast to intrinsic carbon stars, which obtained their enhanced carbon from their internal nucleosynthesis of helium into carbon during the AGB phase.  
We compared the SED of IRAS 09296+1159 with those of dwarf carbon stars and CH giant carbon stars.
In dwarf carbon stars, cool dust emission only becomes prominent long-ward of 12 $\mu$m, while as we see in Figure~\ref{SEDs}, the dust emission around IRAS 09296+1159 begins to become prominent long-ward of 1.2 $\mu$m and is much warmer.
A better match to the SED of IRAS 09296+1159 is that of the CH carbon star HD 100764 (IRAS 11331$-$1418), which we providentially observed spectroscopically as a comparison star for IRAS 09296+1159.
Its SED is also shown for comparison in Figure~\ref{SEDs}.  While the SED of HD 100764 does not have a distinct peak in the mid-infrared, it does have a similarly-broad profile.  This suggests that the dust has a range of temperatures, as is commonly found in a circumstellar disk.  
HD 100764 is known to have hydrocarbon emission features in its mid-infrared spectrum at 6.3, $\sim$8, and 11.3 $\mu$m \citep{sloan07}, and IRAS 09296+1159 also has evidence for emission features at $\sim$8 and 11.3 $\mu$m and a well-established 3.3 $\mu$m feature \citep{geb90}.
 \citet{sloan07} concluded that HD 100764 is a single red giant star with a circumstellar disk which is not greatly obscuring the star.  
 
 The luminosities of IRAS 09296+1159 and HD 100764 were found by integrating the SEDs of the two object.  Both objects are at similarly high galactic latitudes, +41$\arcdeg$ and +44$\arcdeg$, respectively.  Neglecting extinction, fluxes of 1.2 $\times$ 10$^{-12}$ W~m$^{-2}$ (IRAS 09296+1159) and 1.7 $\times$ 10$^{-11}$ W~m$^{-2}$ (HD 100764) were determined.  Using distances determined from Gaia parallaxes \citep{bai18} and assuming optically-thin dust, we determined luminosities of 430 L$_\sun$ (range 300$-$640 L$_\sun$)  at {\it d} = 3.4 kpc for IRAS 09296+1159 and 48$\pm$1 L$_\sun$ at {\it d} = 0.30 kpc for HD 100764, with the quoted ranges based on the uncertainties in the Gaia distances.
 Thus, while the luminosity of IRAS 09296+1159 is higher than that of a main sequence star, it is much lower than that of a post-AGB PPN, which is in the range of 3$-$10 $\times$ 10$^{3}$ L$_{\sun}$.
For its large distance, it has a relatively large {\it Gaia} proper motion, 0.00913$\pm$0.00010$\arcsec$ yr$^{-1}$.  This translates into a high tangential velocity of 145$\pm$30 km~s$^{-1}$.
 We conclude that IRAS 09296+1159 is a giant CH star, and that it received its excess carbon from an AGB carbon star companion and is now surrounded by a carbon-rich disk.  There is no evidence of this companion in the spectra or SED.  Radial velocity monitoring could be undertaken to search for the remnant of this companion.  
 Note that the similar object HD 100764 does not display periodic light variations.  We examined its light curve based on ASAS-3 data and found seasonal variations on the order of 0.1 mag peak-to-peak with no cyclical variability.

The light variations of IRAS 08359$-$1644 differ significantly from that of the other four stars in this study.  Instead of a regular, approximately sinusoidal variation in brightness, it shows sudden, non-periodic drops in brightness of $\sim$0.4 mag.  These dominate the light curve and make it difficult to determine if there is a low-amplitude, periodic light variation at maximum light.
 These sudden drops in brightness are reminiscent of the light drops found in R Coronae Borealis (RCB) stars, although much less extreme, since in RCB stars they can reach several magnitudes.  The light drops in these stars are attributed to a sudden dust formation events that obscure the light from the star \citep{clay96}.
RCB stars are carbon rich and hydrogen poor, and the spectrum of IRAS 08359$-$1644 does not fit this description and is not that of an RCB star.
Nevertheless, we attribute the light drops in IRAS 08359$-$1644 to variable obscuration by dust. 
The reddening ratio at the four deeper drops in light is $\Delta$({\it V$-$R$_C$})/$\Delta${\it V} = 0.16.  
This is equivalent to an extinction ratio of {\it A({\it R$_C$})}/{\it A({\it V})} = 0.84, consistent with dust scattering following an interstellar extinction law \citep{cox00}.

The stellar luminosity of IRAS 08359$-$1644 was similarly determined by integration of the SED and the use of the Gaia distance of 3.6 kpc.  This resulted in a luminosity of 160 L$_\sun$ (range 130$-$200 L$_\sun$), neglecting extinction.  This luminosity is consistent with it being a giant star.
In searching the literature for a giant star with a similar infrared excess, we found a similar SED for the K2~III star HDE 233517 (IRAS 08189+5314; see \citet{reb15}).  
This object has a luminosity of 210 $\pm$20 L$_\sun$ at a distance of 0.86 kpc. 
It is a lithium-enhanced red giant which displayed a low-amplitude variation over one season of study with P = 47.9 days \citep{bal00}.  
Interestingly, \citet{jura06} found it to possess hydrocarbon features at 6.3, 8, and 11.3 $\mu$m on the basis of {\it Spitzer} spectra.  They propose that these features are formed in a flared, orbiting disk surrounding this oxygen-rich star.
It would certainly be interesting to obtain a mid-infrared spectrum of IRAS 08359$-$1644 to see if it also possesses hydrocarbon features.
We earlier suggested that the cause of the light drops was due to variable dust obscuration, either dust ejected along the line of sight or circulating dust transiting across the star.  Each of these seem reasonable.

\section{SUMMARY AND CONCLUSIONS}
\label{summary}

In this paper, we present the results of our ten-year photometric study of five medium-bright ({\it V}=13-15 mag) evolved stars with large mid-infrared excesses.  
We carried out observations at three different observatories, which increased the coverage.  
Complementary, publicly-available ASAS-SN data covering the last three to six years are also available for each, although at lower precision.
All five stars were found to vary in light, with four of them displaying periodic variations.  
Four of them are carbon rich, and it may be that the fifth has some carbon-rich dust observable in its mid-infrared spectrum.   Three of them are well-studied proto-planetary nebulae with F$-$G spectral types.  
The primary results are listed below.  

1. The three PPNe vary in light with periods ranging from 71 to 84 days, with small maximum seasonal variations of only 0.13$-$0.19 mag ({\it V}). 

2. One of the PPNe, IRAS 04296+3429, shows an approximately monotonic increase in brightness, which, when combined with earlier data, amounts to 0.18 mag ({\it V}) over 25 seasons of observing.  Another one, IRAS 06530$-$0213 shows an approximately monotonic increase of 0.05 mag ({\it V}) from 2008$-$2016.  The third, IRAS 23304+6147, shows some variations in the mean seasonal level, but over a smaller range and not monotonically.

3. All three of the PPNe show the light and color to be in phase, with the star faintest when reddest and coolest.  This variation is attributed to the pulsation of the stars.

4.  All three of the PPNe fit the general trend found earlier for a larger sample of ten carbon-rich PPNe of having shorter periods with higher temperatures and having low amplitudes for periods 
$<$ 100 days.

5. All three of the PPNe also possess secondary periods, slightly shorter than the primary periods, with period ratios ({\it P}$_2$/{\it P}$_1$) = 0.84$-$0.93.  These are similar to the values found previously for other PPNe.

6. For the other two objects, IRAS 08359$-$1644 and 09296+1159, we calculate luminosities on the order of 100 L$_\sun$ based on Gaia distances.  
Thus they are not PPNe but giant stars.

7. IRAS 09296+1159 displays a shorter period of 47 days but with a larger maximum amplitude of 0.58 mag.  An examination of its SED shows a very broad mid-infrared peak, usually indicative of dust residing in a circumstellar disk.  A visible-band spectrum of the star indicates that it is a carbon star, which we classify as a CH star.  It is known to possess hydrocarbon emission features.  
Thus, it is not a PPNe, but rather a rare carbon-rich red giant.  
The carbon is likely the result of pollution from a companion star which evolved through an AGB carbon star phase, with much of its mass loss now forming the carbon-rich dust disk.

8. IRAS 08359$-$1644 does not display periodic light variations.  Instead it displays non-periodic drops in brightness of up to 0.4 mag, and is reminiscent of the light curve of an RCB star but with smaller light drops.  We suggest that this is due to dimming by circumstellar dust, either emitted sporadicly or in orbit around the star.  A visible spectrum suggests a G1 III spectral type and not the spectrum of an RCB star.  Its mid-infrared emission is strong, although not as relatively strong compared to the visible as is seen in the other four stars in this study.
It is likely the result of mass loss by a 
companion star.  Based on similarities with the star HD 233517, one might speculate that it too might possess hydrocarbon emission features in its mid-infrared spectrum indicative of the carbon enrichment of a more evolved companion.

In this study of three PPNe, we have added to our knowledge of the pulsation properties of this transient class of post-AGB objects.
The other two, although not PPNe, are interesting in their own right as members of a small group of red giant stars with large mid-infrared excesses caused by circumstellar dust, likely attributed to a more evolved companion which has not been observed.  
Mid-infrared spectroscopic observations of these two red giants would be valuable to more clearly categorize the composition of their circumstellar dust.
Also, useful evidence for a companion could be sought through radial velocity monitoring.

\acknowledgments 
We thankfully acknowledge the many Valparaiso University undergraduate students who carried out the photometric observations from 2008 to 2018 that were used in this study.
These are, in chronological order, Ryan McGuire, Sam Schaub, Chris Wagner, Kristie (Shaw) Nault, Zach Nault, Wesley Cheek, Joel Rogers, Rachael Jensema, Chris Miko, Austin Bain, Hannah Rotter, Aaron Seider, Allyse Appel, Brendan Ferris, Justin Reed, Jacob Bowman, Ryan Braun, Dani Crispo, Stephen Freund,  Chris Morrissey, Cole Hancock, Abigail Vance, Kathryn Willenbrink, Matthew Bremer, Avery Jackson, David Vogl, Tim Bimler, and Sammantha Nowak-Wolff.
We thank K.Volk for integrating the SEDs of the four red giant stars and a helpful discussion of their nature, and T. Robertson for providing some calibration data for the SARA photometry.
We thank the referee and the AAS statistics editor for their helpful comments which led to improvements in the paper.
We acknowledge the assistance of VU undergraduates Kathryn Willenbrink and Michael Battipaglia in the reduction and analysis of the spectra and Justin Reed in the reduction of the photometry. 
The ongoing work of Paul Nord in maintaining the Valparaiso University Observatory is gratefully acknowledged.
BJH acknowledges onging support from the National Science
Foundation (0407087, 1009974, 1413660) and the Indiana Space Grant Consortium.
This research has made use of the SIMBAD database, operated at CDS, Strasbourg,
France, and NASA's Astrophysical Data System.

\section{APPENDIX I: COMBINING THE MULTI-TELESCOPE OBSERVATIONS}
\label{Append-I}

The photometric observations were made with three different telescope, one at the VUO and two that are part of the SARA telescope consortium.  
Photometric observations at the VUO made since June 2008 used an SBIG 6303 CCD camera equipped with Johnson {\it BV} and Cousins {\it R$_C$I$_C$} filters.
The observations made with the SARA-KP and SARA-CT telescopes employed three different detectors each, and in one case, two different filter wheels.  
As described in Section \ref{Obs}, systematic differences in the differential magnitudes were seen between some of the different contemporaneous data sets. 
We attribute the source of these systematic magnitude differences to the neglect of second-order color and extinction coefficients in the transformation of these red stars to the standard system. 
This led us to empirically determine offset values between each of the SARA telescope-detector-filter data sets and the VUO SBIG data set.  
These offset values are listed in Table~\ref{Offsets} for the three stars observed with the SARA telescopes.  Details of the detectors are given by \citet{keel17}.
We estimate the uncertainties in these offset values to be $\pm$0.01 to $\pm$0.02 mag.

\placetable{Offsets} 

The photometric data for the individual stars in this program, with the offsets included, are listed in Table~\ref{diffmag}, 
which is available electronically in a machine readable form. 
These include the time of the observation (HJD), the standardized differential magnitude (program star $-$ comparison star 1), and a code to identify the particular telescope-detector-filter set used.  
These codes are also listed in Table~\ref{Offsets}.

\placetable{diffmag} 

\section{APPENDIX II: RELIABILITY OF THE PERIOD ANALYSES}
\label{Append-II}

\begin{figure}\figurenum{5}\epsscale{1.0} 
\plotone{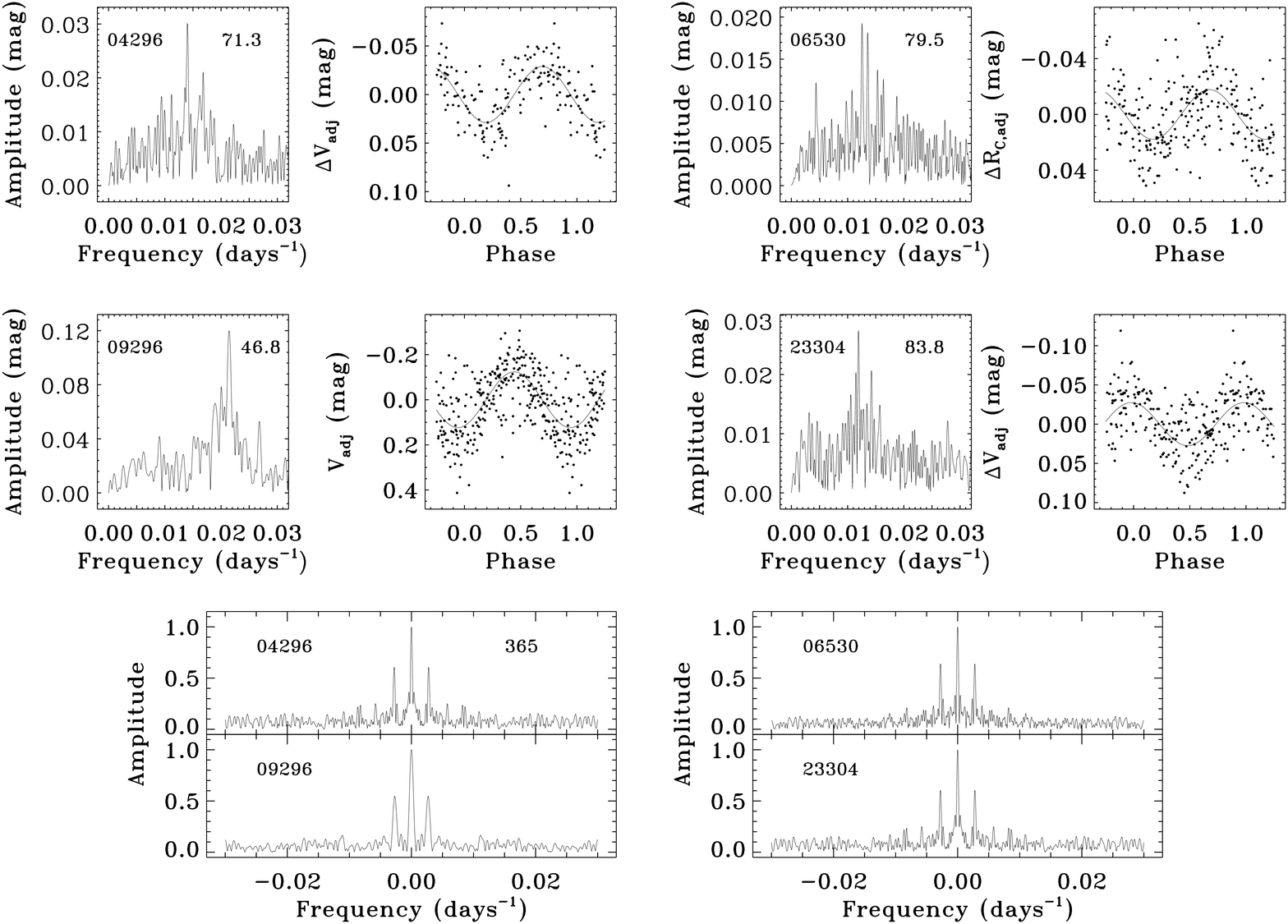}
\caption{ The frequency spectrum and the observations, phased with their dominant periods, for the four periodic objects. The period, in days, is listed in the upper right of each stellar frequency spectrum.  The bottom two panels show the frequency spectrum of the observing window for the four objects, on the same scale as the frequency spectrum.  The side lobes correspond to a period of one year. 
For IRAS 04296+3429 and 06530$-$0213, the trends in the data were first removed and for IRAS 06530$-$0213, 09296+1159, and 23304+6147, the data were seasonally-normalized, as described in the text.
\label{Pspec}}
\epsscale{1.0}
\end{figure}

The primary program used for the determination of the periods of the program objects was PERIOD04.  It is a commonly used program, especially in the astroseismology community.  PERIOD04 uses a discrete Fourier transform algorithm to calculate the frequency spectrum of these unequally-spaced observations.  It has the advantage of allowing one to simultaneously fit several sine curves to the data and determine the best fit by a least-squares analysis.  The program determines the period, amplitude, and phase shift of each sine curve, the zero-point of the brightness, and the least-squares fit uncertainty in each parameter, along with the standard deviation of the sine curve fit to the observations.  
Uncertainties were also computed using a Monte Carlo simulation.  This used a set of data points with the same times as the observed data points, but with magnitudes based on the predicted values from the least-squares fit plus Gaussian noise \citep{lenz05}.
This resulted in uncertainties that were similar to those determined from the least-squares fit uncertainty in each parameter.
The frequency analysis was carried out over the frequency interval from 0.00 to the Nyquist frequency, which was in the range 0.20$-$0.50 days$^{-1}$, with a step size of about 1.5$\times$10$^{-5}$ days$^{-1}$.
The {\it S/N} for each period was calculated by comparing the peak of the Fourier spectrum to the average noise in predefined range about the peak.  For the determination of the noise, we used a frequency range of 0.1 days$^{-1}$ centered on the peak, and we determined the noise after pre-whitening the peak. 
This frequency range provided a good sample of the noise while not being severely impacted by secondary periods, as might occur if we chose a much smaller range.
As stated in Section~\ref{periodogram}, we used the criterion of {\it S/N} $\ge$ 4.0 to indicate a significant result.  
Support for this criterion of significance is found in the study of \citet{kus97}, who were analyzing stellar microvariability using the {\it Hubble Space Telescope} Fine Guidance Sensors observations.  They analyzed a time-series of noise, based on simulated white noise combined with the time series of their observations, and found that, with a Gaussian white noise distribution, a {\it S/N} criterion of 3.6 eliminated 99$\%$ of the noise peaks and a {\it S/N} criterion of 4.0 eliminated 99.9$\%$.
While not claiming white noise in our observations, we do consider this a reasonably conservative criterion to use to distinguish true periods from noise in our data.
Formal uncertainties in the periods were determined from PERIOD04 based on least-squares fittings of the light curves, and these ranged from 0.1 to 0.2 days for these ten-year time intervals.  
The frequency spectrum of the observing window is shown in Figure~\ref{Pspec} for each of the four periodic variables.  One can see in the side lobes a frequency peak at a period of one year that corresponds to the seasonal observing pattern.

As shown in Table~\ref{periods}, for the three objects for which periods were determined from our VUO and SARA observations, 
the resulting periods from the {\it V}, {\it R$_C$}, and ({\it V$-$R$_C$}) light curves are in agreement for each star.  
For the primary periods for the three objects, the {\it S/N} were as follows $-$ IRAS 04296+3429: 7.5 ({\it V}) and 6.8 ({\it R$_C$}), for IRAS 06530$-$0213: 5.4 ({\it V}) and 5.6 ({\it R$_C$}), and IRAS 23304+6247: 5.7 ({\it V}), 4.9 ({\it R$_C$}), and 4.9 ({\it V$-$R$_C$}).   
For IRAS 09296+1159, for which the ASAS-SN data were used in the analysis, the {\it S/N} was 7.4.
Thus, they are clearly well above the commonly accepted significance level.
The secondary periods are also in agreement among the different filters, although at not as high a significance level.

To support the reliability of the results from PERIOD04, we also analyzed the data using several other readily available periodogram programs, and determined similar results.
The additional programs used were the Lomb-Scargle method \citep{sca82}, the CLEANest method \citep{fos95}, and the phase-dispersion-minimization method.  
Analyzing the {\it V} light curve data, we determined the same periods from three programs as we did from PERIOD04 $-$ IRAS 04296+3429: 71.3$-$71.5 vs. 71.3 days, 06530$-$0213: 79.5$-$79.8 vs. 79.5 days, 23304+6147: 83.7$-$83.8 vs. 83.8 days, and 09296+1159: 46.6$-$46.7 vs. 46.7 days.   

Thus we obtained the same period results from PERIOD04 as we did using several other analyses methods. 
We also found good agreement between these periods and those determined from the independent earlier observations from 1994 to 2007 for IRAS 04296+3429 ({\it P}=71 days) and 23304+6147  ({\it P}=85 days). 
The ASAS-SN data sets, over intervals of only 2 to 4 yrs (except in the case of IRAS 09296+1159), show similar results to our new data sets, as discussed earlier (Section~\ref{LC-Var}). 
Taken together, the agreement between the results for the different filters, the agreement between the results of the different analysis methods, and even the general agreement between the results of the different data sets observed a decade apart in time, give good confidence in the reliability of the periods determined.


\clearpage

\tablenum{1}
\begin{deluxetable}{clrrrccccl}
\rotate
\tablecaption{List of PPN Targets Observed \label{object_list}}
\tabletypesize{\footnotesize} \tablewidth{0pt} \tablehead{
\colhead{IRAS ID}&\colhead{GSC ID\tablenotemark{a}}&\colhead{2MASS ID}&\colhead{R.A.\tablenotemark{b}}&\colhead{Decl.\tablenotemark{b}}
&\colhead{{\it l}}&\colhead{{\it b}}
&\colhead{V\tablenotemark{c}}&\colhead{B$-$V\tablenotemark{c}}&\colhead{Sp.T.} \\
&&&(2000.0)&(2000.0)&($\arcdeg$)&($\arcdeg$)&\colhead{(mag)} &\colhead{(mag)} & } 
\startdata
04296+3429   & NCD1000552 & J04325697+3436123     & 04:32:57.0 & +34:36:12   & 166.2 & $-$09.0  & 14.0 & 2.0 & G0~Ia\tablenotemark{d}, F7~I\tablenotemark{e}, F3~I\tablenotemark{f} \\
06530$-$0213 & S3A4020611    & J06553181$-$0217283   & 06:55:31.8 & $-$02:17:28 & 215.4 & $-$00.1  & 14.0 & 2.4 & F5~Ia\tablenotemark{g}, G1~I\tablenotemark{e} \\
08359$-$1644 & S5WN000681    & J08381414$-$1655159   & 08:38:14.1 & $-$16:55:16 & 240.9 & +14.5    & 13.1 & 1.2 & {\bf G1~III}\tablenotemark{i} \\
09296$+$1159 & N8WG000415    & J09322353+1146033    & 09:32:23.5 & +11:46:03   & 221.1 & +41.0    & 13.8 & 1.2 & CH star\tablenotemark{h}, {\bf CH star}\tablenotemark{i} \\
23304+6147   & N1A6000417 & J23324479+6203491     & 23:32:44.8 & +62:03:49   & 113.9 & +00.6    & 13.0 & 2.3 & G2~Ia\tablenotemark{d} \\
\enddata
\tablenotetext{a}{Hubble Space Telescope Guide Star Catalog II (GSC-II), version 2.3.2 (2006).}
\tablenotetext{b}{Coordinates from the 2MASS Catalog.}
\tablenotetext{c}{These values are all variable as discussed in this paper.}
\tablenotetext{d}{Spectral types by \citet{hri95}. }
\tablenotetext{e}{Spectral types by \citet{suarez06}. } 
\tablenotetext{f}{Spectral type by \citet{sancon08}. }
\tablenotetext{g}{Spectral type by \citet{hri03}. }
\tablenotetext{h}{Spectral type by \citet{ji16}. }
\tablenotetext{i}{Spectral type from this study (shown in bold face).}
\end{deluxetable}

\clearpage 
\tablenum{2}
\begin{deluxetable}{lrrc}
\tablecaption{Observed Standard Magnitudes and Colors of the Program Stars
\label{std_ppn}}
\tabletypesize{\footnotesize} 
\tablewidth{0pt} \tablehead{ \colhead{IRAS ID} &\colhead{{\it V}}
 &\colhead{{\it V$-$R$_C$}} 
&\colhead{Date}  \\
&\colhead{(mag)}
 &\colhead{(mag)} 
&\colhead{} } 
\startdata
04296$+$3429  & 14.02   &  1.23 &  2013 Dec 27  \\ 
                          & 14.04   &  1.23 &  2014 Jan 07  \\ 
06530$-$0213   & 14.00  & 1.34 &  2013 Dec 20  \\ 
                           & 14.02  & 1.34 &  2015 Dec 06  \\ 
08359$-$1644  & 13.15   & 0.65 &  2013 Dec 27  \\ 
                         & 13.14   &  0.66 &  2014 Jan 07  \\ 
                           & 13.07 & 0.64 & 2015 Dec 06  \\ 
09296$+$1159\tablenotemark{a}  & \nodata   & \nodata  & \nodata \\ 
23304$+$6147\tablenotemark{b}  & 12.99 & 1.34  & 1995 Oct 17 \\ 
\enddata
\tablecomments{Uncertainties in the brightness and color are $\pm$0.015$-$0.02 mag. }
\tablenotetext{a}{We have not carried out standardized absolute photometry of this object. }
\tablenotetext{b}{Previously published by \citet{hri10}. }
\end{deluxetable}


\tablenum{3}
\begin{deluxetable}{lllrlc}
\tablecaption{Comparison Star Identifications and Observed Standard Magnitudes \label{std_comp}}
\tabletypesize{\footnotesize}
\tablewidth{0pt} \tablehead{ \colhead{IRAS Field}
&\colhead{Object} &\colhead{GSC ID\tablenotemark{a} } &\colhead{2MASS ID} &\colhead{{\it V}} 
&\colhead{V$-$R$_C$}  \\ 
& & & &\colhead{(mag)} &\colhead{(mag)} 
} \startdata
04296$+$3429  & C1 & NCD8000258 & 04331115+3436511 & 13.18  & 0.50   \\
	      & C2 & NCD1000557 & 04323656+3435121 & 13.60 & 1.22   \\
              & C3 & NCD1000565 & 04325697+3436123 & 13.71 & 1.29   \\
06530$-$0213   & C1 & S3A4000501 & 06552133$-$0218580 & 12.59  & 0.05  \\
              & C2 & S3A4000555 & 06553840$-$0222099 & 12.39 & 0.09  \\
              & C3 & S3A4000453 & 06553428$-$0215191 & 12.93 & 0.33   \\
08359$-$1644  & C1 & S5WN000685 & 08380283$-$1655483 & 12.95 & 0.37  \\
              & C2 & S5WM001092 & 08375508$-$1656141 & 13.25 & 0.61 \\
              & C3 & S5WN000696 & 08375695$-$1658133 &13.26 & 0.38  \\
09296$+$1159  & C1 & N8WG000418 & 09323133+1145319 & 13.1\tablenotemark{b} & \nodata \\
              & C2 & N8WG000410 & 09322230+1147599 & 14.4\tablenotemark{b} & \nodata \\
              & C3\tablenotemark{c} & N8WG000420 & 09321208+1145169 & 15.0\tablenotemark{b} & \nodata  \\
23304$+$6147 & C1 & N1A6000418 & 23325467+6203372 & 12.73\tablenotemark{d} & 0.34\tablenotemark{d} \\
              & C2 & N1A6000421 & 23324587+6203175 & 12.73\tablenotemark{d} & 0.45\tablenotemark{d}   \\
              & C3\tablenotemark{e} & N1A6000438 & 23323344+6200569 & 12.44\tablenotemark{d} & 0.46\tablenotemark{d}  \\
\enddata
\tablecomments{Uncertainties in the brightness and color are $\pm$0.01$-$0.02 mag.}
\tablenotetext{a}{Hubble Space Telescope Guide Star Catalog II (GSC-II), version 2.3.2 (2006).}
\tablenotetext{b}{We have not measured the standardized absolute photometry of this object.  Values listed are from \citet{hen12}.}
\tablenotetext{c}{Perhaps some low level variation in C3 within a season, but less than 3$\sigma$. }
\tablenotetext{d}{Previously published by \citet{hri10}. }
\tablenotetext{e}{The brightness of C3 increases by $-$0.02 mag from 2009$-$2018.}
\end{deluxetable}

\clearpage

\tablenum{4}
\begin{deluxetable}{lcrrrrrrrrr}
\tablecolumns{17} \tabletypesize{\scriptsize}
\tablecaption{Periodogram Study of the Light and Color Curves\tablenotemark{a,b}\label{periods}}
\rotate
\tabletypesize{\footnotesize} 
\tablewidth{0pt} \tablehead{ 
\colhead{IRAS ID} &\colhead{Filter} & \colhead{Years}&\colhead{No.} & \colhead{{\it P}$_1$}&\colhead{{\it A}$_1$} &\colhead{{\it $\phi$}$_1$\tablenotemark{c}}&\colhead{{\it P}$_2$}&\colhead{{\it A}$_2$} &\colhead{{\it $\phi$}$_2$\tablenotemark{c}}
&\colhead{$\sigma$\tablenotemark{d}}\\
 & &\colhead{Obs.}  &\colhead{Obs.} & \colhead{(days)}&\colhead{(mag)} & &\colhead{(days)}&\colhead{(mag)}
 &  &\colhead{(mag)}}
\startdata
04296$+$3429 & {\it V}\tablenotemark{e} & 2008-2018& 126 & 71.3 & 0.030 & 0.06 & 63.0 & 0.014 & 0.89 & 0.020 \\
04296$+$3429 & {\it R$_C$}\tablenotemark{e} & 2008-2018& 147 & 71.2 & 0.021 & 0.08 & \nodata & \nodata &  \nodata & 0.018 \\
04296$+$3429 & {\it V$-$R$_C$} & 2008-2018& 118 & 71.4: & 0.008 & 0.05 & \nodata & \nodata & \nodata &  0.013 \\
04296$+$3429  & {\it V}\tablenotemark{f,g} & 2014-2018 & 168 & 66.0 & 0.023 & 0.24 & 73.8 & 0.019 & 0.88 & 0.025 \\ 
04296$+$3429  & {\it V}\tablenotemark{f,g} & 2014-2017 & 124 & 73.7 & 0.029 & 0.85 & 67.6 & 0.023 & 0.88 & 0.023 \\ 
\\
06530$-$0213 & {\it V}\tablenotemark{e,f} & 2008-2018& 174 & 79.5 & 0.019 & 0.07 & 73.9: & 0.014 & 0.00 & 0.023 \\
06530$-$0213  & {\it R$_C$}\tablenotemark{e,f} & 2008-2018& 182 & 79.5 & 0.014 & 0.05 & 73.7 & 0.013 & 0.95  & 0.019 \\
06530$-$0213  & {\it V}\tablenotemark{g} & 2014-2018 & 252 & 73.8 & 0.030 & 0.93 & 78.9 & 0.025 & 0.88 & 0.029 \\ 
\\
09296$+$1159 & {\it V}\tablenotemark{f,g}  & 2012-2018 & 259 & 46.7 & 0.124 & 0.31 & \nodata & \nodata & \nodata & 0.106 \\
09296$+$1159 & {\it V}\tablenotemark{f,g}  & 2012-2018 & 259 & 46.8 & 0.122 & 0.40 & 49.8 & 0.075 & 0.13  & 0.093 \\
\\
23304$+$6147 & {\it V}\tablenotemark{f} & 2009-2018 & 192 & 83.8 & 0.026 & 0.76 & 70.2 & 0.018 & 0.15  & 0.026 \\
23304$+$6147  & {\it R$_C$}\tablenotemark{f} & 2009-2018 & 196 & 83.9 & 0.018 & 0.76 & 70.1: & 0.013 & 0.18 & 0.021 \\
23304$+$6147  & {\it V-R$_C$}\tablenotemark{f} & 2009-2018 & 186 & 84.1 & 0.007 & 0.84 & 70.2 & 0.007 & 0.09 & 0.010 \\
23304$+$6147 & {\it V}\tablenotemark{f,g} & 2015-2017 & 166 & 81.6 & 0.047 & 0.00 & 87.6 & 0.040 & 0.96 & 0.023 \\
\enddata
\tablenotetext{a}{The uncertainties in {\it P}, {\it A}, $\phi$ are approximately $\pm$0.1$-$0.3 day, $\pm$0.002$-$0.003 mag, $\pm$0.02$-$0.03, respectively, with the larger uncertainties in {\it P} found in the shorter time interval ASAS-SN light curves, and the larger uncertainties in $\phi$ found in the smaller amplitude ({\it V$-$R$_C$}) color curves.}
\tablenotetext{b}{Colons (:) indicate less certain period values that fell slightly below our adopted level of significance; see text for details.}
\tablenotetext{c}{The phases are determined based on the epoch of 2,455,600.0000, and they each represent the phase derived from a sine curve fit to the data, not the phase of minimum light.}
\tablenotetext{d}{Standard deviation of the observations from the sine-curve fit.}
\tablenotetext{e}{Analysis based on the light curve with the long-term trend first removed.}
\tablenotetext{f}{Analysis based on the seasonally-normalized light curve.}
\tablenotetext{g}{ASAS-SN data.}
\end{deluxetable}

\clearpage

\tablenum{5}
\begin{deluxetable}{rrrcrcrrl}
\rotate
\tablecaption{Results of Our Period and Light Curve Study\tablenotemark{a} \label{results}} 
\tabletypesize{\footnotesize}
\tablewidth{0pt} \tablehead{ \colhead{IRAS ID} &\colhead{{\it V}} &\colhead{$\Delta${\it V}\tablenotemark{b}} 
&\colhead{SpT} &\colhead{{\it T}$_{\rm eff}$\tablenotemark{c}} &\colhead{{\it P}$_1$} &\colhead{{\it P}$_2$} &\colhead{{\it P}$_2$/{\it P}$_1$} & \colhead{Comments}\\
\colhead{} &\colhead{(mag)} &\colhead{(mag)} &\colhead{} &\colhead{(K)} &\colhead{(days)} &\colhead{(days)} &\colhead{} & \colhead{}}
\startdata
04296$+$3429 & 14.0 & 0.13 &  G0Ia/F7I & 7000 & 71.3 & 63.0 & 0.88 & Trend of increasing brightness \\
06530$-$0213 & 14.0 & 0.18  & F5Ia/G1I & 7310 & 79.5 & 73.7 & 0.93 & Trend of increasing brightness \\
08359$-$1644 & 13.1 & 0.10:\tablenotemark{d} & G1III & \nodata & \nodata & \nodata & \nodata & Drops of $\Delta$V$\sim$0.4; mild RCB-like?\\
09296$+$1159 & 13.8 & 0.58 & C-H1.5 & \nodata & 46.8 & 49.8 & 1.06& Variation in median seasonal values \\
23304$+$6147 &  13.0 & 0.19 & G2Ia & 6750 & 83.8 & 70.2 & 0.84 & \nodata \\
\enddata
\tablenotetext{a}{ASAS-SN variable star identifications and periods $-$ 04296$+$3429: J043256.96+343613.1, {\it P} undetermined; 06530$-$0213: J065531.82$-$021728.3, {\it P}=73.6 days; 08359$-$1644:  J083814.10$-$165515.7, {\it P}=499.0 days; 09296$+$1159: J093223.53$+$114603.2, {\it P}=46.4 days; 23304$+$6147: J233244.57$+$620347.4, {\it P}=80.8 days.}
\tablenotetext{b}{The maximum brightness range observed in a season.}
\tablenotetext{c}{Temperatures from high-resolution spectral observations and abundance analyses: IRAS 04296+3429 and 23304+6147 $-$ \citet{vanwin00}; 06530$-$0213 $-$ average of values from \citet[7250 K]{rey04} and \citet[7375 K]{desmedt16}.}
\tablenotetext{d}{The maximum brightness range observed in a season, outside of the large decreases; approximate value as indicated by colons.}
\end{deluxetable}

\tablenum{6}
\begin{deluxetable}{lclllllllllllllll}
\rotate
\tablecaption{SARA Telescope-Detector-Filter Offsets for Each Star\tablenotemark{a}\label{Offsets}}
\tabletypesize{\footnotesize} \tablewidth{0pt} \tablehead{
\colhead{Telescope-}&\colhead{Code\tablenotemark{b}}&&\multicolumn{2}{c}{IRAS 04296+3429}&&\multicolumn{2}{c}{IRAS 06530$-$0213}&&\multicolumn{2}{c}{IRAS 08359$-$1644}&&\multicolumn{2}{c}{IRAS 09296$+$1159}&&\multicolumn{2}{c}{IRAS 23304$+$6147}  \\
\cline{4-5} \cline{7-8} \cline{10-11}  \cline{13-14} \cline{16-17}
\colhead{Detector}&\colhead{}&&\colhead{{\it V}}&\colhead{{\it R}$_C$}&&\colhead{{\it V}}&\colhead{{\it R}$_C$}&&\colhead{{\it V}}&\colhead{{\it R}$_C$}
&&\colhead{{\it V}}&\colhead{{\it R}$_C$}&&\colhead{{\it V}}&\colhead{{\it R}$_C$}\\
\colhead{}&\colhead{}&&\colhead{(mag)}&\colhead{(mag)}&&\colhead{(mag)}&\colhead{(mag)}&&\colhead{(mag)}&\colhead{(mag)}
&&\colhead{(mag)}&\colhead{(mag)}&&\colhead{(mag)}&\colhead{(mag)}}
\startdata
VUO 	SBIG & A && +0.00 & +0.00 &&+0.00 & +0.00 && +0.00  & +0.00  &&+0.00 & +0.00 && +0.00  & +0.00\\
SARA-KP U42   & B && $-$0.02 & $+$0.03   && $-$0.01 & $+$0.03 && $+$0.02 & $+$0.02 && \nodata  & \nodata  && \nodata   & \nodata \\
SARA-KP U42\tablenotemark{c} & C && $-$0.04 & $+$0.015 && $-$0.10 & $+$0.01 && $+$0.00 & $+$0.015 && \nodata  & \nodata  && \nodata   & \nodata \\
SARA-KP FLI\tablenotemark{d}  & D && $-$0.015 &\nodata   && \nodata  & \nodata  && \nodata   & \nodata  && \nodata  & \nodata  && \nodata   & \nodata \\
SARA-KP ARC  & E && $+$0.05 & $+$0.08 && $+$0.00 & $+$0.055 && $+$0.00 & $+$0.00 && \nodata  & \nodata  && \nodata   & \nodata \\
SARA-CT E6     & F && \nodata &\nodata   && $-$0.10 & $-$0.10 && $+$0.035: & $+$0.035: && \nodata  & \nodata  && \nodata   & \nodata \\
SARA-CT ARC & G && \nodata &\nodata   && $-$0.015 & $+$0.015 && $+$0.015 & $+$0.02 && \nodata  & \nodata  && \nodata   & \nodata \\
SARA-CT FLI   & H && \nodata &\nodata   && $-$0.035 & $+$0.01   && $+$0.00 & $+$0.01 && \nodata  & \nodata  && \nodata   & \nodata \\
\enddata
\tablecomments{Uncertainties are $\pm$0.01 to $\pm$0.02 mag; those with colons (:) are more uncertain.}
\tablenotetext{a}{Offsets as compared to the VUO values, in the sense offset value = mag(VUO) $-$ mag(system).  
Thus these offset values were added to the SARA magnitudes to bring them to the VUO photometric system. }
\tablenotetext{b}{Code represents the coding used in the data tables (Tables 7 to 11) to identify the source of the photometry. }
\tablenotetext{c}{Alternate filter set. }
\tablenotetext{d}{Finger Lakes camera used at SARA-KP for a short time in 2009. }
\end{deluxetable}

\tablenum{7}
\begin{deluxetable}{cccccc}
\tablecaption{Differential Standard Magnitudes \label{diffmag}}
\tabletypesize{\footnotesize} \tablewidth{0pt}
\tablehead{\colhead{IRAS ID} & \colhead{HJD({\it V})} & \colhead{$\Delta${\it V}} & \colhead{HJD({\it R}$_C$)} & \colhead{$\Delta${\it R}$_C$} & \colhead{Code\tablenotemark{a}}\\ \colhead{ } & \colhead{} & \colhead{$\mathrm{(mag)}$} & \colhead{} & \colhead{$\mathrm{(mag)}$} & \colhead{ }}
\startdata
IRAS04296+3429 & 54731.9289 & 0.902 & 54731.9363 & 0.192 & B \\
IRAS04296+3429 &  &  & 54775.6653 & 0.187 & A \\
IRAS04296+3429 & 54792.615 & 0.895 & 54792.5865 & 0.181 & A \\
IRAS04296+3429 & 54829.6863 & 0.951 & 54829.6681 & 0.236 & A \\
IRAS04296+3429 & 54839.6324 & 0.885 & 54839.6309 & 0.196 & B \\
IRAS04296+3429 & 54849.6489 & 0.878 & 54849.6475 & 0.161 & B \\
IRAS04296+3429 & 54865.6368 & 0.886 & 54865.6186 & 0.162 & A \\
IRAS04296+3429 & 54868.638 & 0.892 & 54868.6077 & 0.166 & A \\
IRAS04296+3429 & 54879.6016 & 0.926 & 54879.5822 & 0.195 & A \\
IRAS04296+3429 & 54882.5593 & 0.916 & 54882.5414 & 0.202 & A \\
IRAS04296+3429 &  &  & 54884.6401 & 0.204 & B \\
IRAS04296+3429 & 54886.547 & 0.918 & 54886.5295 & 0.218 & A \\
IRAS04296+3429 & 54893.6443 & 0.957 & 54893.6501 & 0.218 & C 
\enddata
\tablecomments{Table \ref{diffmag} is published in its entirety in the machine-readable format. A portion is shown here for guidance regarding its form and content.}
\tablenotetext{a}{This identifies the source of the photometry, as listed in Table~\ref{Offsets}, and the associated added offset, if any, to bring that observation to the VUO system.}
\end{deluxetable}

\clearpage

\clearpage

\clearpage

\clearpage

\end{document}